\documentclass[twocolumn]{aastex63}

\newcommand{\kms}   {km\,s$^{-1}$}

\usepackage{float}

\newcommand{\met}	  {$-$2.50}	
\newcommand{\metlerr} {0.07} 
\newcommand{\metuerr} {0.07} 

\newcommand{\metdisp}     {0.42} 
\newcommand{\metdisplerr} {0.06} 
\newcommand{\metdispuerr} {0.06} 

\submitjournal{ApJ}

\shorttitle{CaHK MDF of Eri~II}
\shortauthors{Fu et al.}

\begin{document}

\title{Metallicity Distribution Function of the Eridanus~II Ultra-Faint Dwarf Galaxy from Hubble Space Telescope Narrow-band Imaging}

\correspondingauthor{Sal Wanying Fu}
\email{swfu@berkeley.edu}

\author[0000-0003-2990-0830]{Sal Wanying Fu}
\affiliation{Department of Astronomy, University of California, Berkeley, Berkeley, CA, 94720, USA}

\author[0000-0002-6442-6030]{Daniel R. Weisz}
\affiliation{Department of Astronomy, University of California, Berkeley, Berkeley, CA, 94720, USA}

\author{Else Starkenburg}
\affiliation{Kapteyn Astronomical Institute, University of Groningen, Postbus 800, 9700 AV, Groningen, the Netherlands}

\author[0000-0002-1349-202X]{Nicolas Martin}
\affiliation{Universit\'{e} de Strasbourg, Observatoire astronomique de Strasbourg, UMR 7550, F-67000 Strasbourg, France}

\author[0000-0002-4863-8842]{Alexander P. Ji}
\affiliation{Department of Astronomy \& Astrophysics, University of Chicago, 5640 S Ellis Avenue, Chicago, IL 60637, USA}
\affiliation{Kavli Institute for Cosmological Physics, University of Chicago, Chicago, IL 60637, USA}

\author[0000-0002-9820-1219]{Ekta Patel}
\affiliation{Department of Astronomy, University of California, Berkeley, Berkeley, CA, 94720, USA}
\affiliation{Miller Institute for Basic Research in Science, 468 Donner Lab,Berkeley, CA 94720, USA}

\author[0000-0002-9604-343X]{Michael Boylan-Kolchin}
\affiliation{Department of Astronomy, The University of Texas at Austin, 2515 Speedway, Stop C1400, Austin, TX 78712-1205, USA}

\author[0000-0003-1184-8114]{Patrick C\^{o}t\'{e}}
\affiliation{National Research Council of Canada, 
Herzberg Astronomy and Astrophysics Research Centre,
Victoria, BC V9E 2E7, Canada}

\author{Andrew E. Dolphin}
\affiliation{Raytheon Technologies, 1151 E. Hermans Road, Tucson, AZ 85756, USA}
\affiliation{Steward Observatory, University of Arizona, 933 North Cherry Avenue, Tucson, AZ 85721-0065 USA}

\author{Nicolas Longeard}
\affiliation{Laboratoire d’astrophysique, \'{E}cole Polytechnique F\'{e}d\'{e}rale de Lausanne (EPFL), Observatoire, 1290 Versoix, Switzerland}

\author{Mario L. Mateo}
\affiliation{Department of Astronomy, University of Michigan, 311 West Hall, 1085 S. University Avenue, Ann Arbor, MI 48109, USA}

\author[0000-0002-7393-3595]{Nathan R. Sandford}
\affiliation{Department of Astronomy, University of California, Berkeley, Berkeley, CA, 94720, USA}

\begin{abstract}

We use deep narrowband Ca H\&K ($F395N$) imaging taken with the Hubble Space Telescope (HST) to construct the metallicity distribution function (MDF) of Local Group (LG) ultra-faint dwarf (UFD) galaxy Eridanus {\sc II} (Eri {\sc II}). When combined with archival $F475W$ and $F814W$ data, we measure metallicities for 60 resolved red giant branch stars as faint as $m_{F475W}\sim24$ mag, a factor of $\sim4$x more stars than current spectroscopic MDF determinations. We find that Eri {\sc II} has a mean metallicity of [Fe/H]$=$\met$^{+\metuerr}_{-\metlerr}$ and a dispersion of $\sigma_{\mbox{[Fe/H]}}=\metdisp^{+\metdispuerr}_{-\metdisplerr}$, which are consistent with spectroscopic MDFs, though more precisely constrained owing to a larger sample. We identify a handful of extremely metal-poor star candidates (EMP; [Fe/H] $< -3$) that are marginally bright enough for spectroscopic follow up. Eri {\sc II}'s MDF appears well-described by a leaky box chemical evolution model. We also compute an updated orbital history for Eri II using Gaia eDR3 proper motions, and find that it is likely on first infall into the Milky Way. Our findings suggest that Eri II underwent an evolutionary history similar to that of an isolated galaxy. Compared to MDFs for select cosmological simulations of similar mass galaxies, we find that Eri {\sc II} has a lower fraction of stars with [Fe/H] $< -3$, though such comparisons should currently be treated with caution due to a paucity of simulations, selection effects, and known limitations of Ca H\&K for EMPs. This study demonstrates the power of deep HST CaHK imaging for measuring the MDFs of UFDs.

\end{abstract}


\section{Introduction} 
\label{sec:intro}

\par The advent of wide-field photometric surveys at the turn of the 21st century has accelerated the discovery of ultra-faint dwarf galaxies (UFD; e.g., \citealt{willman2005ufd}, \citealt{belokurov2007sdss}, \citealt{bechtol2015DESsat}, \citealt{koposov2015DESsat}, \citealt{laevens2015discovery}) around the Milky Way (MW). These galaxies currently make up the faintest end of the galaxy luminosity function, defined tentatively by \citet{simon2019review} as being fainter than $10^5~L_{\odot}$. There are strong cosmological motivations to study UFDs: their presence constrains the small-scale behavior of dark matter models (e.g., \citealt{bullock2017dmreview}, \citealt{nadler2021}, \citealt{kim2021}), and their ages ($\sim13$~Gyr, e.g., \citealt{brown2014sfh}, \citealt{weisz2014sfh}) make them ideal candidates for being pristine relics from the era of reionization (e.g., \citealt{bovill2009}, \citealt{blandhawthorn2015},  \citealt{weisz2017reionization}). 

\par Additionally, despite having short periods of star formation (SFHs), UFDs with resolved metallicity dispersions display significant internal spreads in stellar metallicity (e.g., \citealt{willman2011wil1},  \citealt{frebel2014segue1}). These observations suggest that they experienced complex chemical enrichment histories that distinguish them from star clusters \citep[e.g., ][]{willman2012galaxydefined}. Any complete theory of galaxy formation must be able to reproduce the properties of this population, and the small sizes of UFDs make them particularly sensitive to the specific implementation of physics in cosmological simulations (e.g., \citealt{munshi2019}, \citealt{agertz2020edge}).

\par Previous studies of the more luminous Local Group (LG) dwarf galaxies have demonstrated that well-populated metallicity distribution functions (MDFs) can be used to trace gas dynamics throughout their star formation histories (e.g., \citealt{tolstoy2009}, \citealt{kirby2011MDFs}, \citealt{kirby2013LZR}, \citealt{ross2015}, \citealt{kirby2017}, \citealt{jenkins2021vlt}). To learn about the physics of galaxy formation at the lowest-known masses to-date, it is of great scientific interest to extend the availability of well-populated MDFs to the lower-luminosity UFDs. However, current spectroscopic studies have struggled to resolve metallicity dispersions in UFD candidates due to very few stars that are bright enough to be efficiently targeted, or observed at all, by current spectrographs (e.g., \citealt{walker2016}, \citealt{martin2016}, \citealt{li2018carinas}, \citealt{simon2020ufds}). Although next-generation photometric surveys such as those from the Rubin Observatory are predicted to find many more UFDs at farther distances (e.g., \citealt{wheeler2019ufdsims}, \citealt{applebaum2021}, \citealt{mutlupatkdil2021}), spectroscopic facilities on future ELT-class telescopes may at best reach down to the sub-giant branch for Segue 1-luminosity UFD galaxies beyond 100~kpc (e.g., Figure 9 from \citealt{simon2019review}). 

\par One well-established alternative approach to measuring stellar metallicities is using medium or narrowband photometry (e.g., \citealt{stromgren1966}, \citealt{mcclure1968photomet}, \citealt{carney1979}, \citealt{geisler1991}, \citealt{lenz1998sdssphotomets}, \citealt{karaali2005sdss}, \citealt{ross2013hstphotomet}).  Photometric filters are designed to target specific features in a star's spectrum that are sensitive to intrinsic properties such as temperature or metallicity.  Thanks to extensive calibration efforts, photometric metallicities are now used expansively to study the Milky Way (e.g., \citealt{helmi2003MW}, \citealt{ivezic2008sdssphotomet}, \citealt{an2013mwhalo}, \citealt{huang2019mwtomography}, \citealt{youakim2020mpmw}, \citealt{arentson2020bulge}, \citealt{cenarro2019jplus}, \citealt{chiti2021skymapper}, \citealt{whitten2021splus}), and can provide accurate metallicities for fainter stars than are accessible through spectroscopy.

\par One particularly useful photometric band for stellar metallicity is the Calcium H\&K (CaHK) doublet in the blue optical at 3968.5 and 3933.7~\AA~(e.g., \citealt{zinn1980}, \citealt{beers1985cahk}, \citealt{anthonytwarog1991}). More recently, the Pristine survey verified the promise for Ca H\&K imaging around the CaHK feature to trace stellar metallicity for FGK stars \citep[e.g. ][]{starkenburg2017pristine} and search the Milky Way for metal-poor stars (e.g., \citealt{youakim2017pristine}, \citealt{aguado2019pristine}, \citealt{venn2020pristine}). Subsequent ground-based imaging studies that leverage similar filters targeting CaHK features in Milky Way satellites (e.g., \citealt{longeard2018dracoII}, \citealt{han2020narrowbandsubaru}, \citealt{chiti2020tucIIskymapper}, \citealt{longeard2021SgrIIGC}) have demonstrated that this technique can substantially expand the sample of stars with metallicity measurements in these systems. 

\par We extend this existing narrowband work in UFDs to even fainter stars and more distant galaxies using the \textit{Hubble Space Telescope} (HST). Specifically, HST-GO-15901 (PI D. Weisz) observed 18 UFD candidates in the F395N filter, which is analogous to the narrow-band CaHK filter used by the Pristine survey. This paper is the first in a series based on these observations.  Here, we demonstrate the power of HST Ca H\&K imaging for uncovering the MDFs of UFDs using new and archival HST imaging of the MW satellite, Eridanus~II (Eri II).

\par Eri~II was initially discovered in Dark Energy Survey data by \citet{bechtol2015DESsat} and \citet{koposov2015DESsat}. Subsequent deeper imaging by \citet{crnojevic2016eriII} and \citep{munoz2018structural} confirm that at  $M_V=-7.1$\footnote{We note that the exact luminosity depends on the adopted distance of Eri~II.}, Eri~II is the faintest known galaxy to host a star cluster. Spectroscopic studies of Eri~II by \citet[][16 stars on Magellan/IMACS]{li2017eriII} and \citet[][26 stars on VLT/MUSE]{zoutendijk2020eriII} confirm that Eri~II is a dark matter-dominated dwarf galaxy ($\sigma_{vel}=6.9^{+1.2}_{-0.9}$~\kms) with a substantial internal spread in metallicity ($\langle \mbox{[Fe/H]} \rangle=-2.38\pm0.13$ and $\sigma_{\mbox{[Fe/H}}=0.47^{+0.12}_{-0.09}$; \citealt{li2017eriII}). Star formation history (SFH) studies of Eri~II, measured from HST broad-band imaging (\citealt{simon2021EriII}, \citealt{gallart2021eriii}, and \citealt{alzate2021eriii}), show that its color-magnitude diagram is consistent with having formed stars in a short burst ($\sim$100~Myr) that ended $\sim$13 Gyr ago. 

\par Eri~II is ideal for demonstrating the efficacy of MDF inference from narrow-band photometry. As the most luminous galaxy in our program, it not only provides a large sample of stars from which to measure a secure MDF, it also has a modest sample of spectroscopic metallcities, which is unusual for most UFDs, allowing us to cross-check our findings.

\par In this paper, we combine HST WFC3/UVIS imaging of Eri~II in the narrow-band $F395N$ filter with broadband archival HST photometry to measure metallicities for 60 stars in Eri~II. In Section \ref{sec:data}, we describe our observations and data reduction process. In Section \ref{sec:metallicity_indiv}, we describe our process for translating our data into metallicity measurements. In Section \ref{sec:results}, we present the resulting MDF of Eri~II, as well as analytic fits. In Section \ref{sec:discussion}, we discuss the our results in the broader context of galaxy formation and cosmology. We conclude with forward-looking remarks in Section \ref{sec:conclusion}.

\section{Data and Observations} 
\label{sec:data}

\subsection{Photometry}

\par We observed Eri~II using HST on February 19, 2020 as part of HST-GO-15901 (PI D. Weisz) using the WFC3/UVIS camera and in the $F395N$ narrow-band filter for two orbits. We performed small dithers to remove hot pixels and reject cosmic rays, choosing patterns that were used to measure proper motions of nearby galaxies from previous Treasury program HST-GO-14734 (PI N. Kallivayalil). In this study, we also used spatially overlapping archival ACS/WFC $F475W$ and $F814W$ broadband imaging of Eri~II, taken as part of HST-GO-14224 (6 orbits; PI C. Gallart, \citealt{gallart2021eriii}).

\par
\begin{figure}
    \centering
    \includegraphics[scale=0.4]{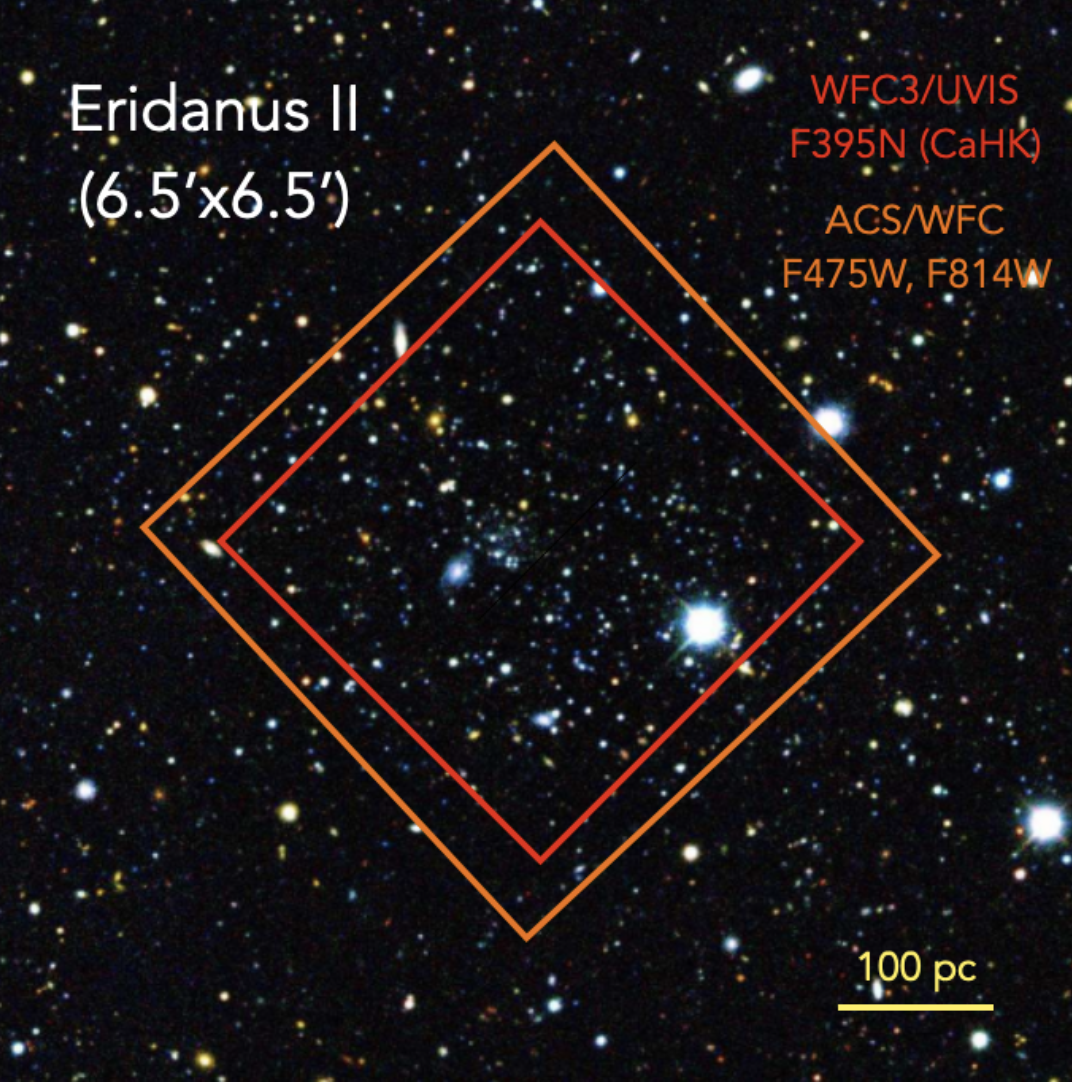}
    \caption{On-sky rendering of Eri~II and our HST frames. }
    \label{fig:spatial}
\end{figure}

 Figure \ref{fig:spatial} shows the placement of our WFC3/UVIS imaging relative to Eri~II and to the footprint of the archival ACS observations. At the distance of Eri~II, the WFC3/UVIS and ACS/WFC imaging subtends $\sim$290 and $\sim$360 pc, respectively, across one side of the image. We also note that both frames capture Eri~II's star cluster, which will analyze the cluster in a future paper. 

\par We use \texttt{DOLPHOT} (\citealt{dolphot2016}, \citealt{hstphot2000}) to perform point-spread function (PSF) photometry simultaneously on the $F395N$, $F475W$, and $F814W$ \texttt{flc} images of Eri~II. From the resulting catalog of objects recovered by \texttt{DOLPHOT}, we select stars by applying a quality cut to the deep broadband imaging. Specifically, we only include stars with $\mbox{S/N} > 5$, $|\mbox{sharp}|^2 < 0.3$ and $\mbox{crowd} < 1$ in each band. 

\par
\begin{figure*}
    \epsscale{1.2}
    \centering
    \includegraphics[scale=0.6]{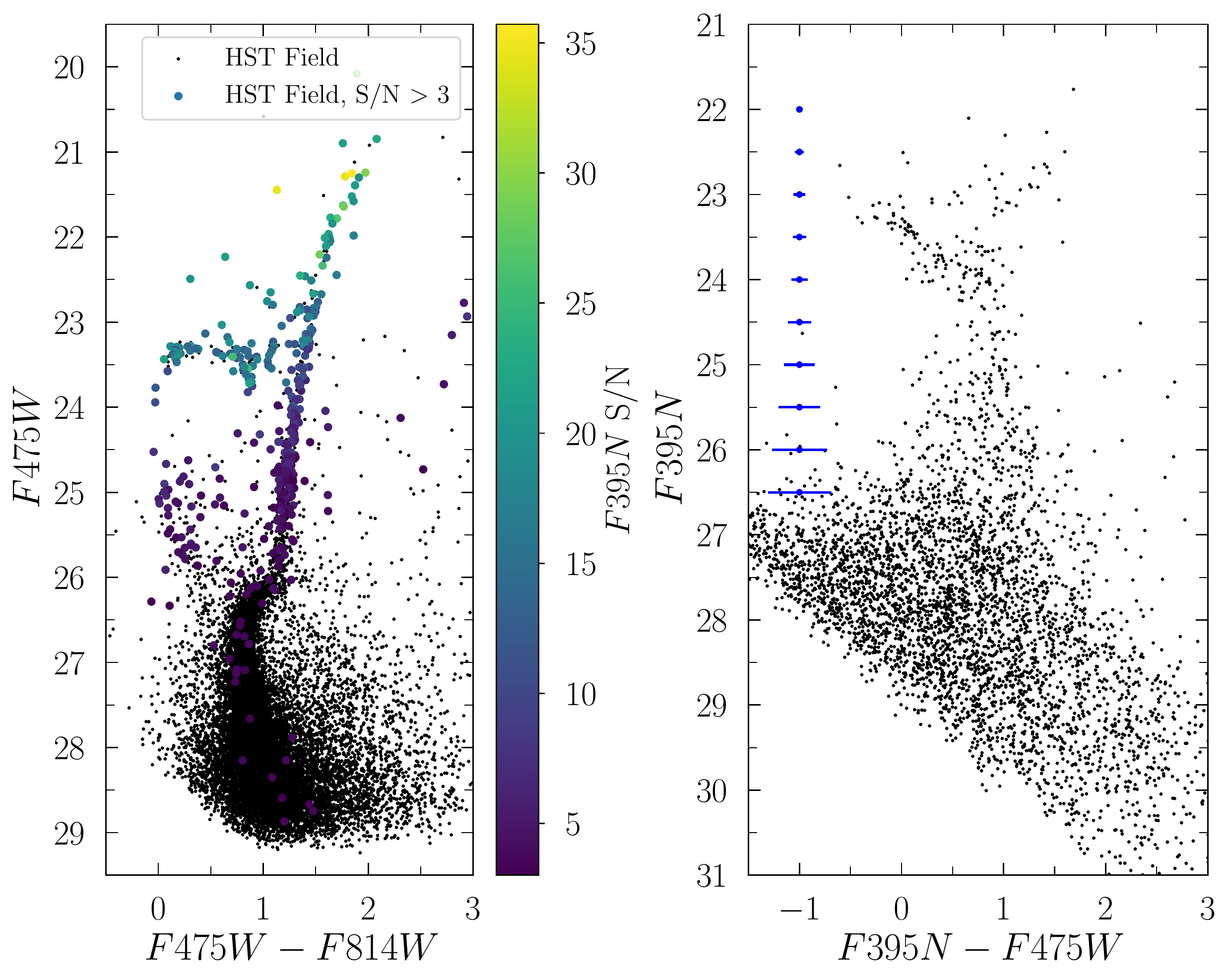}
    \caption{\textbf{Left:} The broadband CMD of the Eri~II field population for stars that pass the quality cuts. We color-code points with their corresponding S/N in F395N for S/N $>3$. In this panel and the next, the photometry plotted excludes stars within two half-light radii of the galaxy's cluster. \textbf{Right:} CMD of the Eri~II field population with F395N. The blue error bars show the typical uncertainty as a function of F395N. Uncertainties on $F395N - F475W$ color are dominated by F395N.}
    \label{fig:EriII_CMD}
\end{figure*}

 Figure \ref{fig:EriII_CMD} shows the broadband color-magnitude diagram (CMD) of stars that pass the quality cuts in each filter. Color-coded stars in the left panel are those with $F395N$ S/N $>$ 3. The right panel shows the $F395N$ narrow-band CMD. The blue error bars show the typical uncertainty in color as a function of $F395N$. 

\subsection{Selection of Stars for Photometric Metallicity Determination} 

\par We apply further cuts to select stars for our metallicity inference in Section \ref{sec:metallicity_indiv}, following the convention of past studies that use narrow-band photometry of the Ca H\&K lines to study the MDF of Milky Way satellites. To begin, we select stars with $F395N$ S/N $>$ 10: this is the S/N threshold above which the photometric measurements can provide reliable metallicities (e.g., \citealt{longeard2018dracoII}). We also remove any stars that fall within two half-light radii of the galaxy's star cluster using the structural parameters provided by \citealt{simon2021EriII} ($r_h \sim 0.16\arcmin$), as the focus of this paper is on the field population. From the resulting sample, we limit our analysis to only stars that fall along the red giant branch (RGB) of Eri~II. We exclude horizontal branch stars in the galaxy that also have sufficient S/N in $F395N$ because the wavelength region of the narrow-band filter is dominated by the Balmer lines in these hot stars, rather than the metallicity sensitive H\&K lines \citep[e.g., ][]{starkenburg2017pristine}.

\par The left panel of Figure \ref{fig:eriII_CaHK} presents the final sample of stars that we use for our analysis. The orange box on the broadband CMD indicates the box used to select the final sample of stars. These 60 stars have $F395N$ S/N ranging from $\sim10$ to 36. We verify that the star at $F475W\sim20.8$ and $F475W-F814W\sim2.0$, which falls right above the selection box, is brighter than the TRGB of a metal-poor isochrone shifted to the distance modulus of Eri~II from \citet{crnojevic2016eriII} and reported in Table \ref{tab:EriIIprop}, and therefore exclude it from our analysis.

\par The center panel of Figure \ref{fig:eriII_CaHK} shows where our sample falls on the SDSS-Pristine color space defined by the Pristine survey, and which was derived from a similar color space defined by \citet{keller2007skymapper} for the Skymapper survey. \citet{starkenburg2017pristine} demonstrated that the color space $ (g_{SDSS} - i_{SDSS})$\footnote{Technically, \citet{starkenburg2017pristine} apply the extinction correction to their data, whereas we apply extinction to the model as described in Section \ref{sec:metallicity_indiv}.} vs. $(CaHK - g_{SDSS}) - 1.5\,(g_{SDSS} - i_{SDSS})$ was effective for separating stars of different metallicities. For our study, we use the equivalent HST filters $F395N$, $F475W$, $F814W$ to construct an analogous color space where the $x$ axis is defined as $F395N - F475W$, and the $y$ axis of this space is defined as $(F395N - F475W) - 1.5(F475W - F814W)$. We henceforth refer to the $y$ axis color as the CaHK color index, and to the entire color space as the CaHK color space. As shown in the center panel of Figure \ref{fig:eriII_CaHK}, the synthetic HST photometry from MIST isochrones  (\citealt{choi2016mist}, \citealt{dotter2016mist}) also separate in this space according to metallicity. We discuss the data, relative to these models, in Section \ref{sec:metallicity_indiv}.

\subsection{Artificial Star Tests}

\par
\begin{figure*}
    \epsscale{1.2}
    \centering
    \includegraphics[scale=0.33]{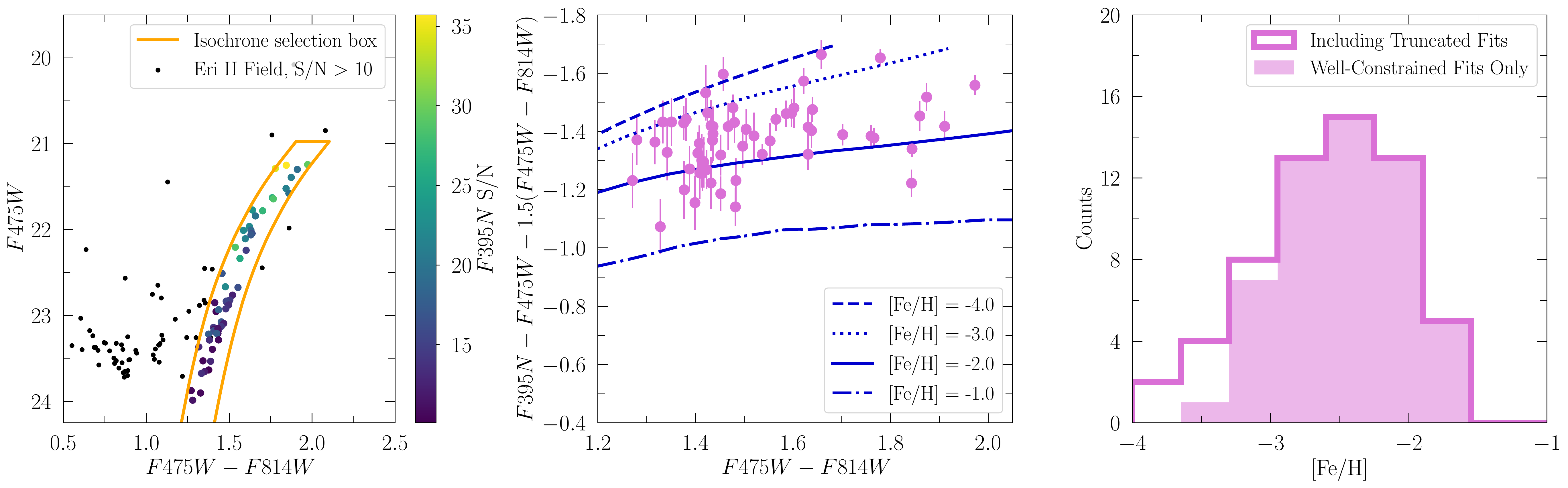}
    \caption{\textbf{Left:} Broadband CMD of Eri~II. Black points are photometry with $F395N$ S/N $>10$. The orange outline is the box that we used to select RGB stars in Eri~II for our MDF analysis, and the color-coded points are the 60 selected stars. Our final sample reaches past the horizontal branch of Eri~II. \textbf{Center:} The sample of Eri~II stars that we use for our MDF analysis, plotted in the CaHK photometry color space defined by \citet{starkenburg2017pristine}. We also apply extinction in the line-of-sight of Eri~II to the MIST mono-metallic isochrone tracks for 13 Gyr stellar populations with $[\alpha/\mbox{Fe}]=+0.4$. We plot these tracks here to demonstrate how they separate in this particular color space. \textbf{Right:} MDF of Eri~II, constructed from point estimates. Binsizes are the median uncertainty of Eri II metallicity measurements (0.35 dex). The outlined histogram includes point metallicity estimates of stars whose posterior distributions describe an upper limit on metallicity due either to model limitations or S/N, which we elaborate on in Section \ref{sec:results}.}
    \label{fig:eriII_CaHK}
\end{figure*}

 We use artificial star tests (ASTs) to compute the uncertainties on our photometry. The procedure for an AST is as follows: we insert an artificial star with input magnitudes in $F395N$, $F475W$, and $F814W$ onto a random position in the corresponding HST \texttt{flc} image. We only input stars whose broadband photometry fall within the isochrone selection box in the left panel of Figure \ref{fig:eriII_CaHK}, and with $F395N$ photometry that fall within the CaHK color space in the middle panel of Figure \ref{fig:eriII_CaHK}. 

\par We then attempt to recover the star in all of these photometric bands using the same PSF-fitting procedure and the same culling properties for our photometric measurements. Additionally, we require that ASTs are within 0.75 mag of their input magnitudes, which helps mitigate spurious blends. The output of an AST is the difference between the measured magnitude and the input magnitude (out-in). We perform 85,580 ASTs. This large sample allows us to quantify (i) average offsets in measured photometry (bias) and (ii) the variance of out-in (error) at all locations in CaHK space. In Section \ref{sec:metallicity_indiv}, we describe how we apply the resulting error profile from ASTs to our metallicity inference.

\begin{deluxetable}{cc}
\tablecaption{Properties of Eri~II}
\tablehead{\colhead{Parameter} & \colhead{Eri~II} } 
\startdata
R.A. (h:m:s)                & 03:44:20.1$\pm$10.5\arcsec   \\ 
Dec. (d:m:s)                & $-$43:32:0.1$\pm$5.3\arcsec  \\ 
Distance modulus (mag)      & 22.8 $\pm$ 0.1               \\ 
Distance (kpc)              & 366  $\pm$ 17                \\ 
Absolute Magnitude ($M_V$)  & $-$7.1                       \\ 
Luminosity ($L_{\odot}$)    &  $6\times10^4$\\
Eccentricity ($\epsilon$)   & 0.48 $\pm$ 0.04              \\ 
Half-light radius (\arcmin) & 2.3 $\pm$ 0.12               \\ 
Half-light radius (pc)      & 277 $\pm$ 14                 \\ 
\enddata
\tablecomments{Structural parameters of Eri~II, derived from \citet{crnojevic2016eriII}.}
\label{tab:EriIIprop}
\end{deluxetable}

\section{Metallicity Measurements}
\label{sec:metallicity_indiv}

\par The middle panel of Figure \ref{fig:eriII_CaHK} shows the location of our 60 Eri~II field RGB stars on the CaHK color-color plot. We overplot the alpha-enhanced ([$\alpha\mbox{/Fe]}=+0.40$), 13~Gyr MIST isochrones for giant stars. The models have been extinction corrected using the dust maps of \citet{sfd1998} and bolometric extinction corrections provided with the MIST models \citep{choi2016mist} which enable us to calculate the extinction for the $F395N$ filter. We also consult the TRILEGAL Milky Way model \citep{vanhollebeke2009trilegal} to determine the extent to which contaminants would affect our sample: due to the small field of view and our particular CMD cuts, the chance that our sample contains contaminants are low. Moreover, the CaHK photometry of potential contaminants imply that they would likely be more metal-rich than $\mbox{[Fe/H]}>-1.0$, which we do not observe in our data. Finally, we also verify that none of the stars in our sample have been deemed Eri~II non-members through spectroscopy (\citealt{li2017eriII}, \citealt{zoutendijk2020eriII}). 

\par The majority of our sample lie between the $\mbox{[Fe/H]}=-3$ and $\mbox{[Fe/H]}=-1$ tracks, with a mean value somewhere between $\mbox{[Fe/H]}=-3$ and $\mbox{[Fe/H]}=-2$. Even by eye, it is apparent that the stars in Eri~II span a range of $\sim2$ dex in metallicity. This observation is robust for both the bright stars and the faint stars in our sample. 

\par We derive metallicities for each star by first retrieving synthetic HST photometry for mono-metallic MIST isochrones; we use the MIST models for our analysis because it is the only currently-available set of evolutionary models that include stellar populations as metal-poor as $\mbox{[Fe/H]}=-4.0$\footnote{We recognize that there are known issues with the performance of MIST/MESA isochrones for metal-poor stars (see, e.g., \citealt{kielty2021}, and references therein), and that refining the models of metal-poor stars remains an active area of research \citep[e.g., ][]{karovicova2020}. However, as we show in Section \ref{sec:results}, the metallicities we derive using MIST are consistent with other Eri~II metallicity measurements in the literature. We deem this result sufficient for this paper in demonstrating the power of CaHK photometry to recover broad features of Eri~II's MDF.}. We limit our analysis to using tracks only for a 13 Gyr stellar population, in accordance with star formation histories (SFH) of Eri~II from broadband HST data (i.e., \citealt{simon2021EriII}, \citealt{gallart2021eriii}, and \citealt{alzate2021eriii}). We obtained advance access to the MIST v2 models, which are the alpha-enhanced version of the MIST v1 models. We use these models, instead of Solar-scale alpha models, to fit Eri~II, because UFDs with short bursts of star formation are generally expected to have enhanced alpha element abundances over a large range of their stellar metallicities (e.g., \citealt{vargas2013alpha}, \citealt{frebel2014segue1}). We refer readers to the Appendix for additional discussion on the impact of alpha enhancements.

\par Next, we apply the results of the ASTs to each mono-metallic model. For each mono-metallic isochrone, we match each model isochrone point to its closest set of AST input magnitudes. From those ASTs, we calculate the bias and apply it to each isochrone point as offsets. We find for the ASTs that the CaHK color index has a positive color bias that increases for lower S/N points. A star with $F475W - F814W = 1.4$ and a CaHK color index of $-1.0$, corresponding to a point on the lower RGB of Eri~II, would have a color bias of $+0.1$ in CaHK. In contrast, a star with $F475W - F814W = 1.8$ and a CaHK color index of $-1.3$, corresponding to the upper RGB of Eri~II, would have a color bias of $+0.025$ in CaHK. As a result, all of the isochrone tracks are shifted redward in CaHK space, though based on the spacing between the tracks and the S/N of our data, they do not result in drastically different metallicities. This bias effect is driven by the shallower $F395N$ observations, and may arise from charge transfer efficiency corrections and/or post-flash effects on the photometry that cause us to lose signal for fainter stars. We intend to investigate this issue in greater detail in future work.

\par After applying the bias from the ASTs, we apply extinction using the dustmaps of \citet{sfd1998} and bolometric corrections from the MIST models \citep{choi2016mist}. Once we apply the instrumental error profile and extinction models to the synthetic photometry, we linearly interpolate the CaHK color index as a function of $F475W-F814W$ to allow for the evaluation of the model at any color value.

\par As the middle panel of Figure \ref{fig:eriII_CaHK} shows, the uncertainty in $F475W - F814W$ color of each star is small compared to the uncertainty in the CaHK color index. While the largest uncertainty in $F475W - F814W$ is $\sim 0.01$ mag, the uncertainty in the CaHK color index is at least $0.03$ mag for the highest S/N points, and $0.1$ mag for the lowest S/N. For simplicity in our fitting, we assume that the error in the $F475W - F814W$ color is negligibly small. We verify that scatter from the ASTs at a given CaHK location are statistically identical to the photometric errors reported by \texttt{DOLPHOT}. For simplicity, we adopt the photometric errors reported by \texttt{DOLPHOT} for our analysis.

\par For each star, we hold its $F475W - F814W$ color as constant and compute the corresponding model CaHK photometry at every metallicity using the tracks interpolated from MIST. To measure the metallicity of an individual star, we adopt a Gaussian likelihood function of the form: 

\begin{equation}
    \mbox{log }L = -\frac{1}{2} \frac{(CaHK - CaHK_m(\mbox{[Fe/H]}))^2}{(\sigma_{CaHK} ^ 2)}
\end{equation}

where $CaHK$ and $\sigma_{CaHK}$ are the corresponding CaHK color index measurement and measurement uncertainty, and $CaHK_m(\mbox{[Fe/H]})$ is the model CaHK color index corresponding to a particular metallicity at a fixed $F475W - F814W$ color. We adopt uniform priors on metallicity over the range of the grid that includes the star's $F475W-F814W$ color. We sample the resulting posterior distribution using \texttt{emcee} \citep{emcee} by initializing 50 walkers and running the MCMC chain for 10000 steps, with a burn-in time of on average 50 steps per star. We monitor for convergence using the Gelman-Rubin statistic \citep{gelmanrubin1992}.

\par We present sample metallicity fits for individual stars in the Appendix and the table of metallicity measurements in Table \ref{tab:indiv_measurements}. 

\section{Results}
\label{sec:results}

\subsection{Individual Measurements}
\label{sec:resultsindiv}

\begin{figure}
    \centering
    \includegraphics[scale=0.35]{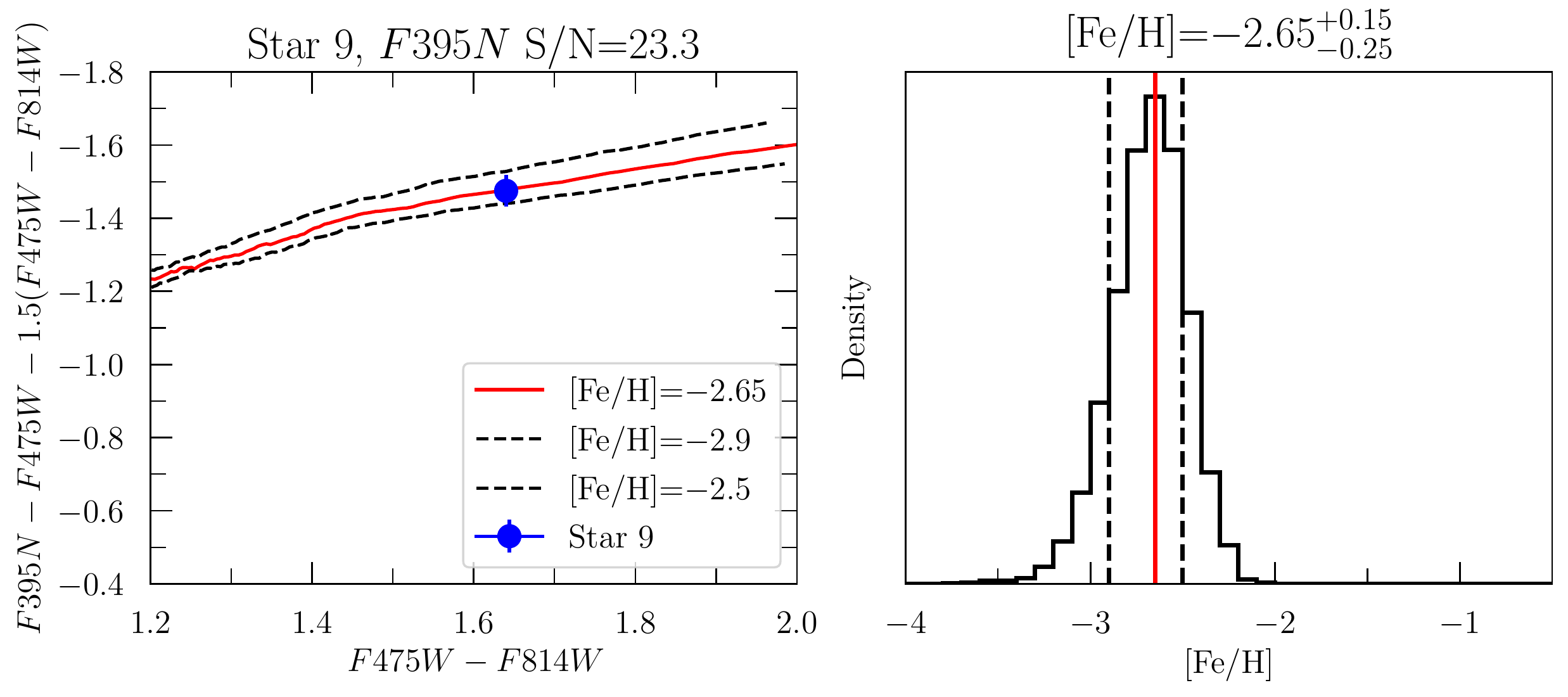}
    \caption{Example of a star (Star 9) with a well-constrained metallicity posterior distribution. \textbf{(left)} The location of Star 9 in CaHK color space, plotted with the MIST isochrone models corresponding to the median, 16th, and 84th percentiles of its metallicity posterior distribution. \textbf{(right)} The metallicity posterior distribution for Star 9.}
    \label{fig:sample_posterior_main}
\end{figure}

\begin{figure*}
    \epsscale{1.2}
    \centering
    \includegraphics[scale=0.5]{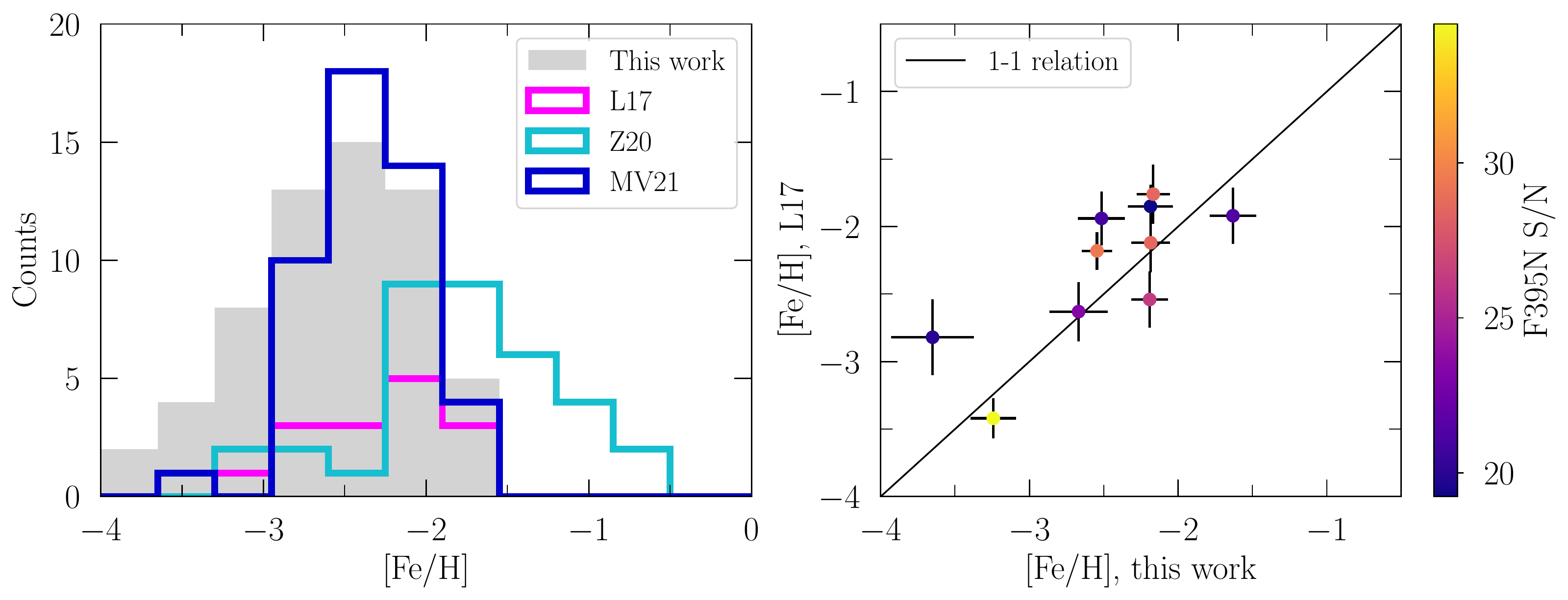}
    \caption{Results of fitting the metallicities of individual stars to CaHK photometry. \textbf{Left:} Histogram of point measurements from this work, compared against those from L17, Z20, and MV21. Binsizes are the median uncertainty of Eri II metallicity measurements (0.35 dex). \textbf{Right:} 1-1 comparison of our measurements against those from L17 for the 10 stars common to both of our samples.}
    \label{fig:indiv_histograms}
\end{figure*}

\par In the right panel of Figure \ref{fig:eriII_CaHK}, we present the MDF of Eri II.  The reported metallicities are the median of the marginalized posterior distribution for each star.  Table \ref{tab:indiv_measurements} lists the median and 68\% confidence interval for each star.  Metallicity uncertainties are typically on the order of 0.1 dex, though they vary depending on the SNR, metallicity of the star, and location on the RGB, as we discuss below.

\par Our MDF of Eri II has a peak between $-$2.5 and $-$2.0. It spans a range of at least 2.0~dex: as we discuss below, some of the most metal-poor stars are consistent with metallicities lower than what our model grid allows. We also present our measurements in Table \ref{tab:indiv_measurements}, and now discuss features in the posterior distributions of our metallicity measurements. 

\par Figure \ref{fig:sample_posterior_main} presents an example of a star with a well-constrained metallicity posterior distribution. A well-defined metallicity posterior distribution has a clear peak and clear tails, and are narrower for higher S/N stars. Metallicity posterior distributions also tend to have longer metal-poor tails because the CaHK models are less distinguishable at the metal-poor end (e.g., middle panel of Figure \ref{fig:eriII_CaHK}). We present additional examples in Figure \ref{fig:normal_posteriors} in the Appendix. 

\par There are 3 stars for which we derive only upper limits on the metallicity. These are stars with high S/N in $F395N$ (S/N$>$15) and whose photometry places them at the edge or beyond the grid of models that we use to make our metallicity inference. We flag these stars as candidate extremely metal-poor stars (EMP; $\mbox{[Fe/H]}<-3.0$) that may be of interest for future spectroscopic studies, and discuss them in more detail in Section \ref{sec:noteworthystars}. We also present the posterior distributions of these stars in Figure \ref{fig:mp_posteriors}.

\par There are 10 stars in our sample whose photometric error bars fall within 1 sigma of the metal-poor grid edge, but whose metal-poor posterior distribution truncates within 2 sigma; we also note these stars in Table \ref{tab:indiv_measurements}. It is possible that at least some of these stars are more metal-poor than the value that we report, and that the measurements we derive here are limited by the models that we use. We provide posterior distributions of these stars in Figure \ref{fig:truncated_posteriors}. In the same vein, there are 4 stars whose 1 sigma error bars are at the edge of our grid, meaning that their posterior distributions show a clear peak, but the metal-poor tail is even more truncated.

\par Finally, there are 4 stars for which we report only an upper limit because of their low S/N photometry (S/N$\sim$10). These are stars whose posterior distributions show no clear peak.

\subsection{Comparison to Literature}

\par In this section, we compare our MDF of Eri~II to those currently in the literature: \citet{li2017eriII}, \citet{zoutendijk2020eriII}, \citet{gallart2021eriii}, and \citet{martinezvazquez2021}, henceforth L17, Z20, G21, and MV21, respectively. LZ17 and Z20 target RGB members of Eri~II using spectroscopy and obtain metallicity measurements for some of the same stars in our sample. From these studies, we can directly compare individual measurements using different methods to evaluate the efficacy of CaHK metallicities. In contrast, for G21 and MV21, we cannot do star-by-star comparisons, but we can compare the overall MDF derived from the various methods. G21 uses deep HST photometry to measure the SFH of Eri~II, inferring an MDF in the process. MV21 studies variable stars in Eri~II, and measures the galaxy's MDF from its RR Lyrae stars.

\par We first discuss how our metallicity measurements compare against the spectroscopic studies. For context, L17 targeted candidate Eri~II members within $8\arcmin$ of the galaxy using Magellan/IMACS spectroscopy. They report metallicities derived using the calcium triplet (CaT) equivalent width calibration \citep{carrera2013cat} for 16 RGB members of Eri~II, down to a magnitude of $g_{DES} \sim 21.7$ mag. Z20 targeted candidate Eri~II members within $1\arcmin$ centered on Eri~II using VLT/MUSE spectroscopy. They report metallicities derived using full-spectrum fitting with the PHOENIX model spectra \citep{husser2013phoenix} for 26 Eri~II member stars, down to a magnitude of $F606W \sim 23.8$ mag.

\par In Figure \ref{fig:indiv_histograms}, we compare our measurements against those of L17 and Z20. The histogram in the left panel compares our MDF with those from L17 and Z20. The right panel compares our metallicity measurements with the 10 common stars we have with the L17 sample. 

\par We find good agreement between our metallicites and those in L17: while the scatter is a little larger than expected from the reported uncertainties, there is no evidence of a systematic offset between the two studies. In contrast, we find disagreement with the results of Z20.  Z20 find systematically more metal-rich stars and report systematically smaller uncertainties.  The MDF of Z20 implies that Eri II would be a more metal-rich system than implied by the dwarf galaxy luminosity-metallicity relationship (LZ; see Section \ref{sec:discussion}). Z20 noted a similar disagreement with L17, but did not explore the origin of this tension. Resolving this issue is beyond the scope of this paper. 

\par Next, we compare our results against those from the photometric studies. G21 studied the SFH of Eri~II from deep HST broadband data and inferred an MDF for the galaxy as part of their work. While their MDF spans the same metallicity range as ours, the shape is qualitatively different from the one that both we and MV21 (below) derive. Given the differences in the technique and the part of the CMD used to infer the metallicities, it is challenging to discern the source of discrepancy.

\par MV21 studied RR Lyrae (RRL) stars in Eri~II using multi-epoch $g$, $r$, and $i$ imaging on the Goodman and Dark Energy Cameras, as well as $F475W$, $F606W$, $F814W$ from HST/ACS. They derive metallicities from 46 RRL stars using the period-luminosity relations from \citet{caceres2008} and \citet{marconi2015}.

\par The left panel of Figure \ref{fig:indiv_histograms} compares our MDF against those derived from the RRL sample in MV21. The peak of our MDF is consistent with that of MV21, though we are able to recover more metal-poor stars. This discrepancy may be expected from using different populations of stars to trace the MDF: MV21 point out that stars in the metal-poor tail of the MDF may not fall into the instability strip as RR Lyrae stars. As a result, MV21 also recover a smaller metallicity dispersion ($\sigma_{\mbox{[Fe/H]}}=0.3$~dex) compared to what we derive in Section \ref{sec:resultsgauss} ($\sigma_{\mbox{[Fe/H]}}=\metdisp$~dex).

\par Overall, these comparisons suggest that our metallicity measurements can recover results that are consistent with those from the literature.

\subsection{Gaussian MDF Fit}
\label{sec:resultsgauss}

\par Following long-standing practice in spectroscopic studies of UFDs (e.g., \citealt{simon2020ufds}, \citealt{li2018carinas}, \citealt{kirby2015ufdspectra}, \citealt{willman2012galaxydefined}), we fit a Gaussian to Eri~II's MDF\footnote{Although, we note that studies such as \citet{leaman2012nongauss} suggest that the MDFs of dwarf galaxies are not well-represented by a Gaussian form. We will explore this further in a study of the full galaxy sample.}. In particular, we use the two-parameter Gaussian likelihood function used by L17, and which was adapted from \citet{walker2006}:

\begin{eqnarray}
\mbox{log}L = -\frac{1}{2} \sum^{N}_{i=1}\mbox{log}(\sigma^2_{\mbox{[Fe/H]}} +\sigma^2_{\mbox{[Fe/H]},i}) \nonumber \\ - \frac{1}{2}\sum^{N}_{i=1}\frac{(\mbox{[Fe/H]}_i - \langle \mbox{[Fe/H]} \rangle)^2}{\sigma^2_{\mbox{[Fe/H]}} +\sigma^2_{\mbox{[Fe/H]},i}}
\end{eqnarray}

\par where $\langle \mbox{[Fe/H]} \rangle$ and $\sigma_{\mbox{[Fe/H]}}$ are the mean metallicity and metallicity dispersion of Eri~II, and $\mbox{[Fe/H]}_i$ and $\sigma_{\mbox{[Fe/H]},i}$ are the metallicity and metallicity uncertainties for each star. We assume uniform priors on the mean, with the maximum and minimum set by the range of the point metallicity measurements, and require that the metallicity dispersion be greater than zero ($\sigma_{\mbox{[Fe/H]}} > 0$). We use \texttt{emcee} to sample the posterior distribution, initializing 50 walkers for 10000 steps. The autocorrelation time for this run is 30 steps. The GR statistic indicates that the chains are converged.

\par For stars with different lower and upper uncertainties measured from the 16th and 84th percentiles of their PDFs, we adopt the average of the error bars to use in our fit. For stars that have only upper limit constraints, we include them in our fit using the medians of their PDFs as the point estimate and 16th/84th percentile values to compute uncertainties. All of these are approximations, and in principle, a more rigorous approach would be to use individual PDFs in our MDF inference. We will consider such an approach in subsequent papers in this series.

\par From 60 stars, we measure a mean metallicity of \met$^{+\metuerr}_{-\metlerr}$ dex and metallicity dispersion of \metdisp$^{+\metdispuerr}_{-\metdisplerr}$ dex. We present the resulting joint distribution in the Appendix as Figure \ref{fig:gaussian_corner}, and tabulate these values in Table \ref{tab:mdf_params}. 

\par In comparison, L17 measure a mean metallicity of $-2.38 \pm 0.13$~dex and a metallicity dispersion of $0.43^{+0.12}_{-0.09}$. Thus, using a completely independent approach to measuring the MDF, we find consistency with and improved precision over the CaT MDF. Beyond enabling a precise measurement of Eri II's MDF, this finding also demonstrates the power of CaHK imaging in practice.

\subsection{Analytic Chemical Evolution Models}
\label{sec:resultsanalytic}

\par The MDF of a galaxy is shaped by the gas dynamics throughout the course of its star formation history. One zone chemical evolution models are simple analytic MDFs that have long been used for interpreting the formation process of various Milky Way components (e.g., \citealt{schmidt1963}, \citealt{lynden-bell1975}, \citealt{pagelpatchett1975}, \citealt{hartwick1976}, \citealt{tinsley1980}), as well as those of its nearby dwarf galaxies (e.g., \citealt{helmi2006}, \citealt{kirby2013LZR}, \citealt{ross2015}). 

\par Prior to this work, the only UFD whose MDF was sufficiently populated enough to fit one-zone chemical evolution models is Bootes~I, which has been analyzed by \citet{jenkins2021vlt}, \citet{romano2015} and \citet{lai2011}. In this section, we follow precedent set by the aforementioned studies by fitting the Leaky Box, Pre-Enriched, and Accretion/Extra Gas chemical evolution models to the MDF of Eri~II. 

\par The Leaky Box model defined by \citet{pagel1997} describes the scenario where a galaxy forms stars from gas that is initially devoid of metals, and loses gas in the process of successive star formation and enrichment. Its single parameter is $p_{\rm eff}$, the effective yield, which encapsulates contributions from both supernova enrichment and gas outflow to the stellar MDF. 

\par The Pre-Enriched model and the Accretion/Extra Gas models are more complex versions of the Leaky Box. The Pre-Enriched model assumes that star formation starts from pre-enriched gas of metallicity $\mbox{[Fe/H]}_0$. As a result, it imposes a floor on the lowest metallicity stars in the galaxy, and simplifies to the Leaky Box in the limit of $\mbox{[Fe/H]}_0 \to -\infty$. The Accretion/Extra Gas model of \citet{lynden-bell1975} describes a system that is allowed to accrete pristine gas during the process of star formation. This model adds an extra parameter, $M$, which is the final stellar mass in terms of the initial gas mass at the onset of star formation. An Accretion model with $M>1$ implies that the galaxy experienced accretion during its star formation history, resulting in an MDF with a larger peak and a smaller metal-poor tail. When $M=1$, the Accretion model reduces to the Leaky Box.

We find the best-fit parameters for these analytic MDFs following \citet{jenkins2021vlt,ji2021ant2cra2}\footnote{Code available at \url{https://github.com/alexji/mdfmodels}}.
Briefly, we adopt the Gaussian likelihood from \citet{kirby2011MDFs} and \citet{kirby2013LZR}: i.e., the likelihood for each star $i$ is the convolution between the MDF and a Gaussian with standard deviation corresponding to that star's metallicity uncertainty $\sigma_{\rm{[Fe/H]},i}$ evaluated at the observed metallicity $\mbox{[Fe/H]}_i$. We treat error bars in our sample the same way as for the Gaussian MDF fit described in the previous section.

\par We use dynamic nested sampling with \texttt{dynesty} \citep{dynesty} to sample the parameters' posterior distribution. We use fairly wide priors: log-uniform for $p_{\rm{eff}}$ from $10^{-3}$ to $10^{-1.1}$; uniform in $\mbox{[Fe/H]}_0$ from $-4$ to $-2$ for the Pre-Enriched model; and uniform in $M$ from 1 to 30 for the Accretion model.

\par We present the posterior distributions in the Appendix as Figures \ref{fig:leakybox_posterior} (Leaky Box), \ref{fig:preenriched_posterior} (Pre-Enriched), and \ref{fig:extragas_posterior} (Accretion), and tabulate fit parameters in Table \ref{tab:mdf_params}. We report the median of the posterior distributions for each parameter, and compute error bars using the 16th and 84th percentiles. For the pre-enriched model, the posterior distribution of $\mbox{[Fe/H]}_0$ is truncated at the metal-poor end. For the accretion model, the posterior distribution of $M$ shows that our data can constrain an upper limit on $M$, but cannot rule out $M=1$ with high confidence. 

\par
\begin{figure}
    \epsscale{1.2}
    \includegraphics[scale=0.4]{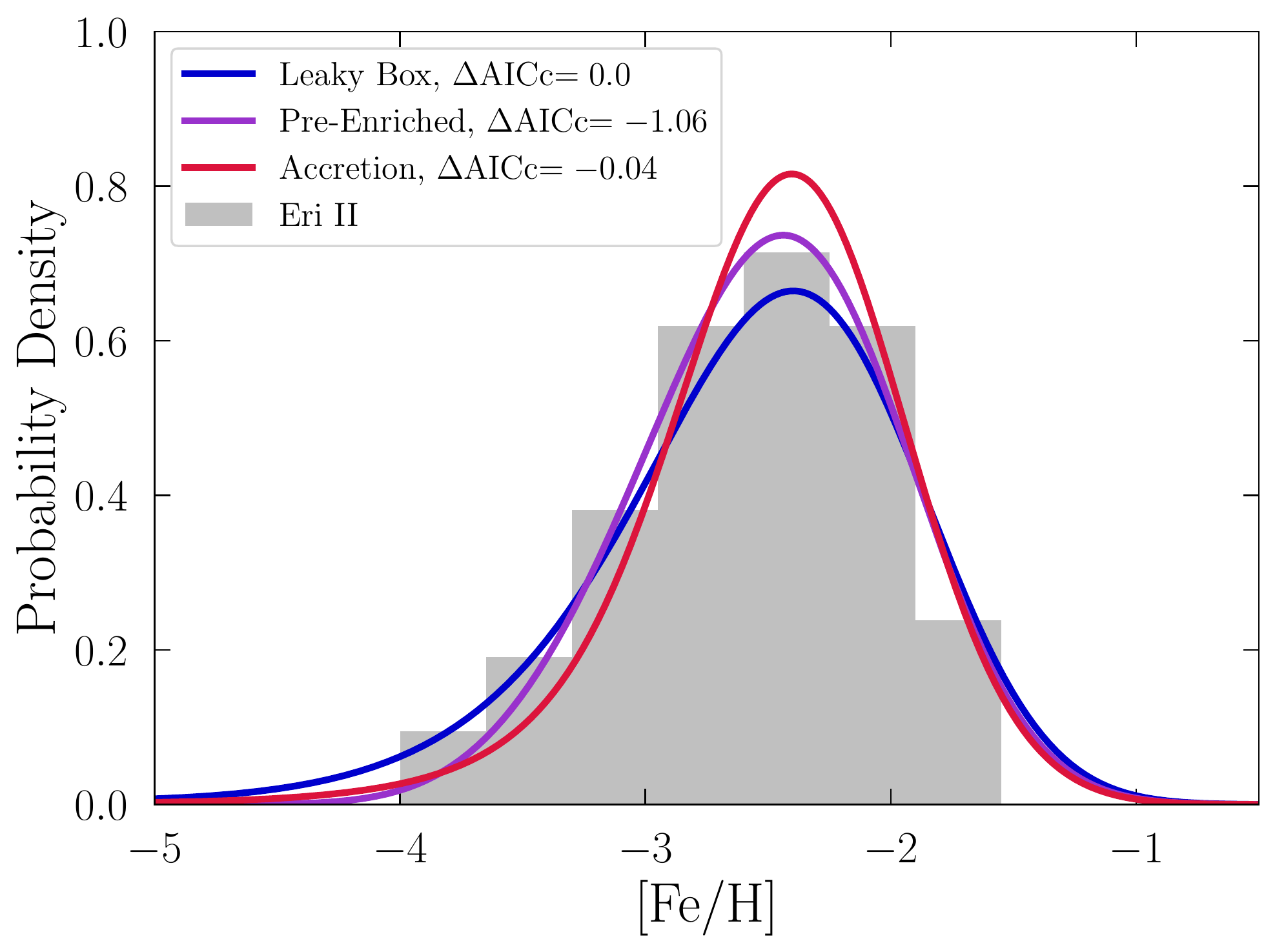}
    \caption{Comparing histogram of Eri~II measurements to best-fit one zone chemical evolution models. Binsizes are the median uncertainty of Eri~II metallicity measurements (0.35~dex), and the one zone models have also been convolved with a Gaussian of the same width.}
    \label{fig:final_histogram}
\end{figure}

 We present the best-fit MDFs in Figure \ref{fig:final_histogram}. By eye, it looks that all models fit the MDF of Eri~II reasonably well, and so we investigate whether the Eri~II data prefer an MDF more complex than a Leaky Box. One way to quantify the model comparison is by using the corrected Akaike Information Criterion \citep[AICc][]{kirby2011MDFs}, which is a penalized likelihood ratio test. We compute this statistic for the Leaky Box, Pre-Enriched, and Accretion models. We tabulate the AICc for the Leaky Box in Table \ref{tab:mdf_params}, and for the other two models, their difference in AICc $\Delta \mbox{AICc}$ from that of the Leaky Box. A positive value for $\Delta \mbox{AICc}$ would suggest that the model is preferred over the Leaky Box, but we find that there is no strong preference between the three models that we fit since the $\Delta \mbox{AICc}$ is small. \texttt{dynesty} also computes the Bayesian evidence, which we choose not use here as it is sensitive to the exact prior volume choice, but it provides similar conclusions as the AICc. 

\par We now discuss limitations in our methodology that could change the analytic chemical evolution model interpretation. As we lay out in the previous subsection, we approximate the posterior distribution of measurements for individual stellar metallicities as Gaussian, despite posterior distributions having longer metal-poor tails due to the indistinguishability of CaHK metallicities in metal-poor regimes. As a result, our current methodology downweighs the metal-poor tail of Eri~II's MDF. Fitting analytic models using the full PDFs of individual stellar metallicities would remedy this issue, and could bring the MDF of Eri~II closer to the Leaky Box model.

\par Additionally, we caution that the one-zone models used in this work oversimplify the problem by assuming that galaxies quench via turning the last of its available gas into stars. This assumption is likely incorrect because star formation is rarely that efficient, and low-mass galaxies can be quenched via reionization, ram pressure stripping, and/or SNe feedback (see \citealt{eggen2021} for example of a $10^5~L_{\odot}$ galaxy in the field at $z=0$ that has been observed to experience SNe feedback, and also the discussion in Section \ref{sec:discussion}). While one way to remedy this issue in future work would be to fit models that truncate star formation at a particular metallicity (see, e.g., the ram pressure stripping model from \citealt{kirby2013LZR}), additional detailed modeling would be fruitful for developing intuition around interpreting various components of a UFD's MDF.

\par With the above caveats and the minimal differences in AICc between the three models in mind, we also note that the results of our Pre-Enriched and Accretion fits suggest that they are also consistent with the Leaky Box limit. We therefore suggest that within the scope of our analysis in this work, the MDF of Eri~II is best represented by the Leaky Box model, and discuss the implications in Section \ref{sec:discussion}.

\begin{deluxetable}{ccc}
\label{tab:mdf_params}
\tablecaption{Parameters of analytic MDF fits to Eri~II for the Gaussian, Leaky Box, Pre-Enriched Gas, and Accretion models. We report the AICc for the Leaky Box, and for the Pre-Enriched Gas and Accretion models, we report the difference in AICc between those models and that of the Leaky Box.}
\tablehead{\colhead{Model} & \colhead{Parameters} & \colhead{Values}}
\startdata
Gaussian & [Fe/H] & \met$^{+\metuerr}_{-\metlerr}$ \\
 & $\sigma_{\mbox{[Fe/H]}}$ & \metdisp$^{+\metdispuerr}_{-\metdisplerr}$ \\
\tableline
Leaky Box & $p_{\mbox{eff}}$ & $0.005^{+0.001}_{-0.001}$ \\
& AICc & 96.04 \\
\tableline 
Pre-Enriched & $p_{\mbox{eff}}$ & $0.004^{+0.001}_{-0.001}$ \\
& $\mbox{[Fe/H]}_0$ & $-3.73^{+0.21}_{-0.17}$ \\
& $\Delta$AICc & $-1.06$ \\
\tableline 
Accretion & $p_{\mbox{eff}}$ & $0.005^{+0.001}_{-0.001}$ \\
& $M$ & $2.77^{+2.59}_{-1.14}$ \\
& $\Delta$AICc & $-0.04$ \\
\enddata
\end{deluxetable}

\subsection{Noteworthy Individual Stars}
\label{sec:noteworthystars}

\par Direct descendants of the first stars (Pop~III) are hypothesized to be more easily identifiable in UFDs, owing to these systems' comparatively simpler enrichment histories in contrast to those of their more luminous counterparts (e.g., \citealt{ji2015popIII}). The chemical abundances of EMPs in UFDs are of great scientific interest because they are thought to trace directly back to past rare chemical enrichment events (e.g. \citealt{ji2016rproc}) or to supernovae (SNe) ejecta from Pop III stars (e.g., \citealt{frebel2015nfcreview}, \citealt{jeon2017popIII}). In this section, we discuss the pathfinding potential for CaHK photometry to contribute to the search for these stars. 

\par
\begin{figure*}
    \epsscale{1.2}
    \centering
    \includegraphics[scale=0.33]{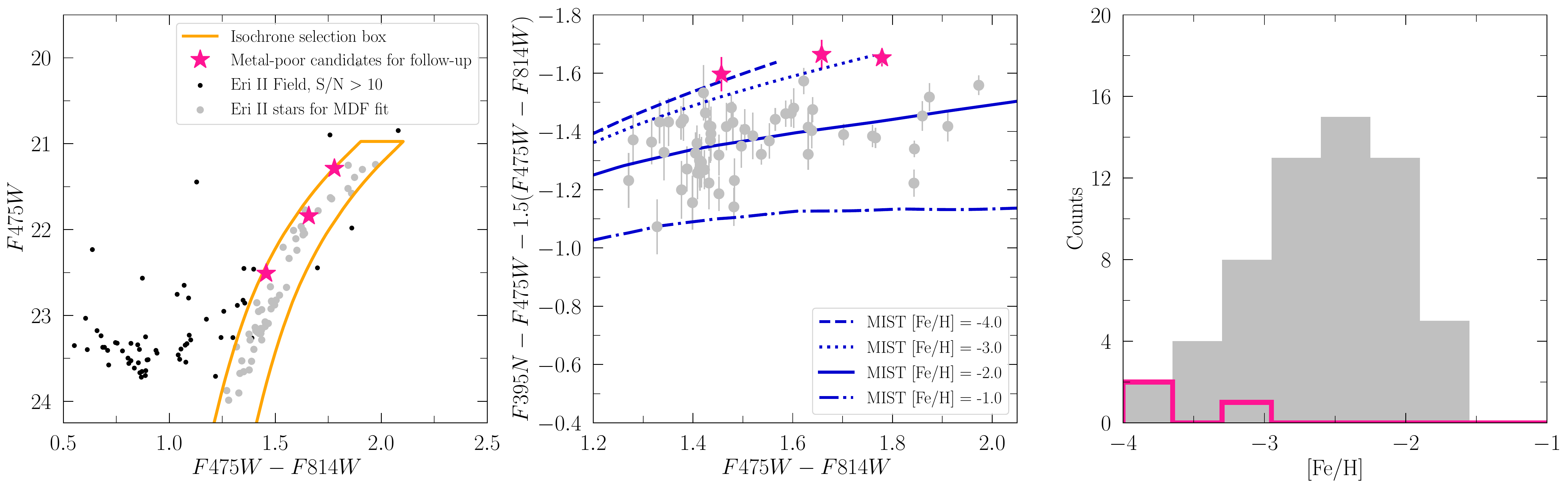}
    \caption{\textbf{Left:} CMD of Eri~II RGB stars, highlighting the metal-poor star candidates (blue) that we discuss in Section \ref{sec:noteworthystars} as being potential candidates for spectroscopic follow-up. \textbf{Center:} Location of these stars in CaHK color space. \textbf{Right:} Histogram demonstrating the contribution that these stars make to the MDF of Eri~II. Binsizes are the median uncertainty of Eri II metallicity measurements (0.35 dex).}
    \label{fig:eriII_outliers}
\end{figure*}

 In Figure \ref{fig:eriII_outliers}, we highlight spectroscopically accessible stars that are at the metal-poor edge of our fitting grid, that may be particularly enlightening for spectroscopic follow-up observations. In particular, we identify 3 stars that are candidate EMPs. As shown in Figure \ref{fig:mp_posteriors}, these stars have truncated metal-poor posteriors distributions, so they may be more metal-poor than our grids allow. They are also located on the upper RGB of Eri~II, with the faintest star being $m_{F475W}\sim22.5$~mag. Although their CMD position on the blue edge of the RGB raises the possibility that they are asymptotic red giant branch (AGB) stars, the purely ancient stellar population of Eri~II ($\sim13$~Gyr) suggests that it is unlikely to have many, or any, AGB stars.

\par Of these 3 stars, two of them have CaT metallicities measured by L17. For star 2 ($m_{F475W}=21.3$, $F395N~\mbox{S/N}=34.5$), we find that it has $\mbox{[Fe/H]}=-3.50\pm0.15$, compared to the L17 measurement of $\mbox{[Fe/H]}=-3.42\pm0.15$: these measurements are in good agreement with each other. For star 11, we obtain a 95\% upper limit of $\mbox{[Fe/H]}=-3.15$, compared to the L17 metallicity of $\mbox{[Fe/H]}=-2.82\pm0.28$. For these two stars, both our measurements and measurements from the literature affirm that they are consistent with having $\mbox{[Fe/H]}<-3.0$, demonstrating the potential for CaHK to identify EMP candidates. Additional spectroscopic observations could aim to ascertain the behavior of $\alpha$ element and C abundances in Eri~II at low metallicities, which in turn can be used to infer the properties of the SNe whose yields contributed to these patterns ($\alpha$, \citealt{nagasawa2018horI}, \citealt{ji2020carinas}; C, \citealt{jeon2021cemp}).

\par Unlike the other two EMP stars that we identify, Star 21 ($m_{F475W}=22.5$, $F395N~\mbox{S/N}=17$) does not have spectroscopic data. We estimate the upper limit of its metallicity to be $\mbox{[Fe/H]}=-3.05$. Additional observations to measure metallicities from CaT spectroscopy or full spectral fitting would be fruitful both for verifying that it is an EMP, as well as measuring its other chemical abundances. 

\subsection{Searching for a Metallicity Gradient in Eri~II}

\par Radial metallicity gradients have long been observed in Local Group dSph galaxies (e.g., \citealt{held1999}, \citealt{saviane2000}, \citealt{saviane2001}, \citealt{harbeck2001}, \citealt{tolstoy2004}, \citealt{deboer2012}, \citealt{kacharov2017}). The physics responsible for the origin and steepness of these gradients remain an active area of research, with proposed mechanisms such as natal angular momentum \citep[e.g., ][]{schroyen2011}, stellar feedback \citep[e.g., ][]{elbadry2016}, and star formation histories \citep[e.g., ][]{mercado2021gradients}.  To-date, however, studies of metallicity gradients in dwarf galaxies have mostly been limited to those more luminous than $10^5~L_{\odot}$ (e.g., \citealt{vargas2014}, \citealt{ho2015}).

\par Motivated by recent searches for metallicity gradients in UFDs (\citealt{chiti2021tucII}, \citealt{longeard2021booI}), we search for a spatial trend in our data. We find none. This could be because: (1) There is none, (2) Our imaging, which covers $<1~r_h$ of Eri~II, is too limited in spatial extent to find one; and/or (3) the uncertainties on the metallicity measurements need to be smaller, if the gradient is weak but non-zero. 

\section{Discussion}
\label{sec:discussion}

\subsection{Improved MDF Statistics for Eri~II}

\par
\begin{figure}
    \epsscale{1.2}
    \includegraphics[scale=0.53]{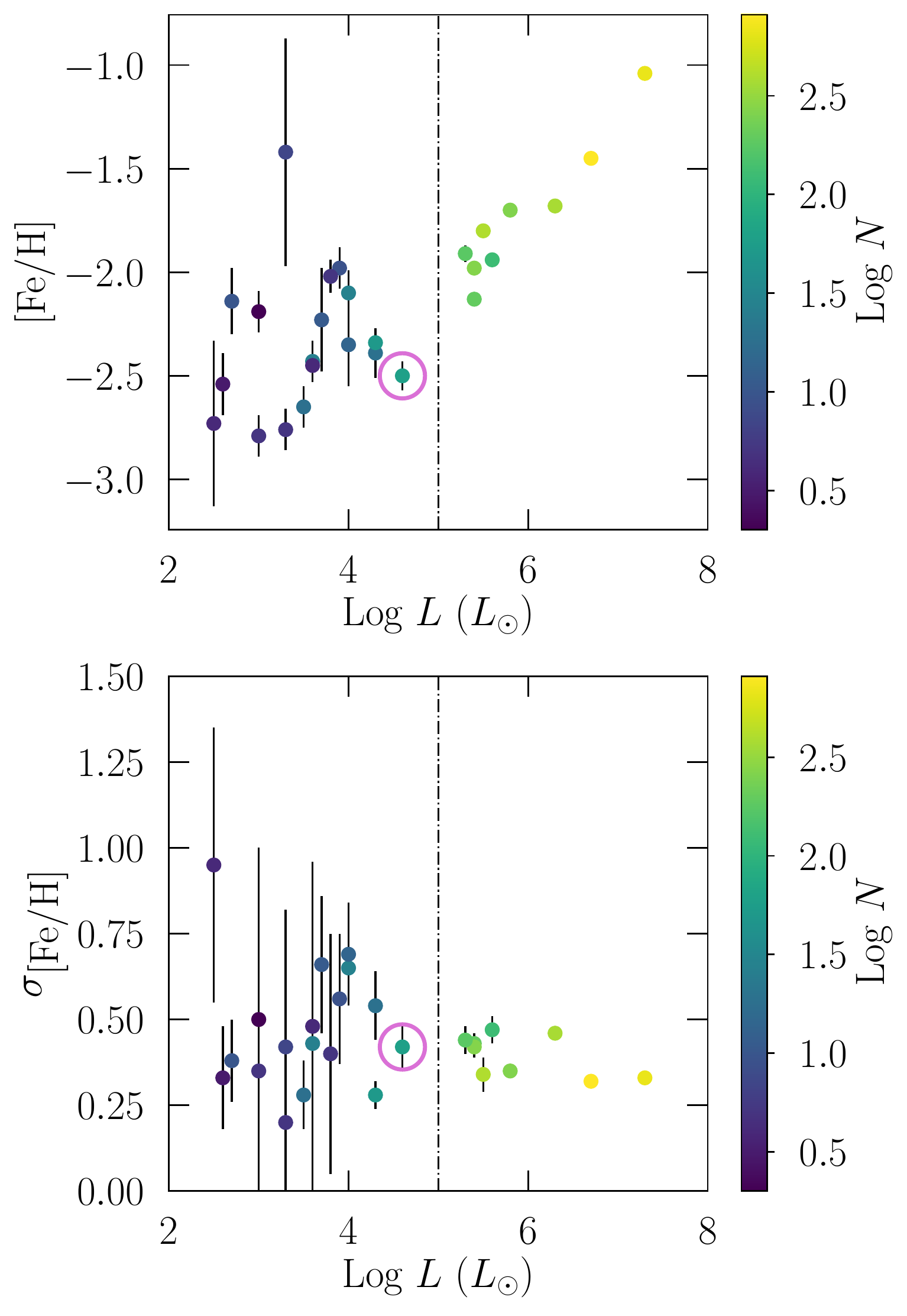}
    \caption{Subset of mean metallicity (top) and mean metallicity dispersions (bottom) for known Milky Way satellites, color coded by number of stars used to make the measurement (compiled partly from \citet{simon2019review}, with updated measurements for Boo~I from \citet{jenkins2021vlt} and including our Eri~II measurement). The dash-dotted line separates the ultra-faint dwarfs from the classical dwarfs. We indicate Eri~II with the purple circle.}
    \label{fig:LZR}
\end{figure}

 Figure \ref{fig:LZR} compares the mean and sigma from fitting a Gaussian to our MDF of Eri II  with those derived for other LG dwarf galaxies (data compiled by \citealt{simon2019review}, with updated measurements for Boo~I from \citealt{jenkins2021vlt}), where the data points are color-coded by the number of stars used to measure the MDF in each system. The vertical line in both panels delineates the separation between UFDs and the classical dwarf galaxies as defined by \citet{simon2019review}. The top panel shows that the mean metallicity we derive for Eri~II is fully consistent with its expected location along the mass-metallicity relationship for dwarf galaxies \citep{kirby2013LZR}. The bottom panel shows that the metallicity dispersion we derive is on par with that of more luminous dwarf spheroidal galaxies, and within the range of dispersions observed so far in LG satellites. The agreement between our Eri II results and the broader dwarf galaxy MDF provides further support for the power of Ca HK-based metallicities.

\par As discussed in Section \ref{sec:resultsgauss}, our MDF fit of Eri~II provides a twofold increase in precision in metallicity dispersion over those of the measurements from L17 due to the expanded sample of stars with metallicity measurements. Out of all the UFDs with measured MDFs so far, Eri~II is also now better-characterized for the same reason: MDF measurements for other UFDs have been limited due to few stars that have been accessible through spectroscopic studies. These results affirm the potential for CaHK imaging to deliver improved UFD MDF statistics. 

\subsection{Quenching Mechanisms for Eri~II}

\begin{figure}
    \centering
    \includegraphics[scale=0.55]{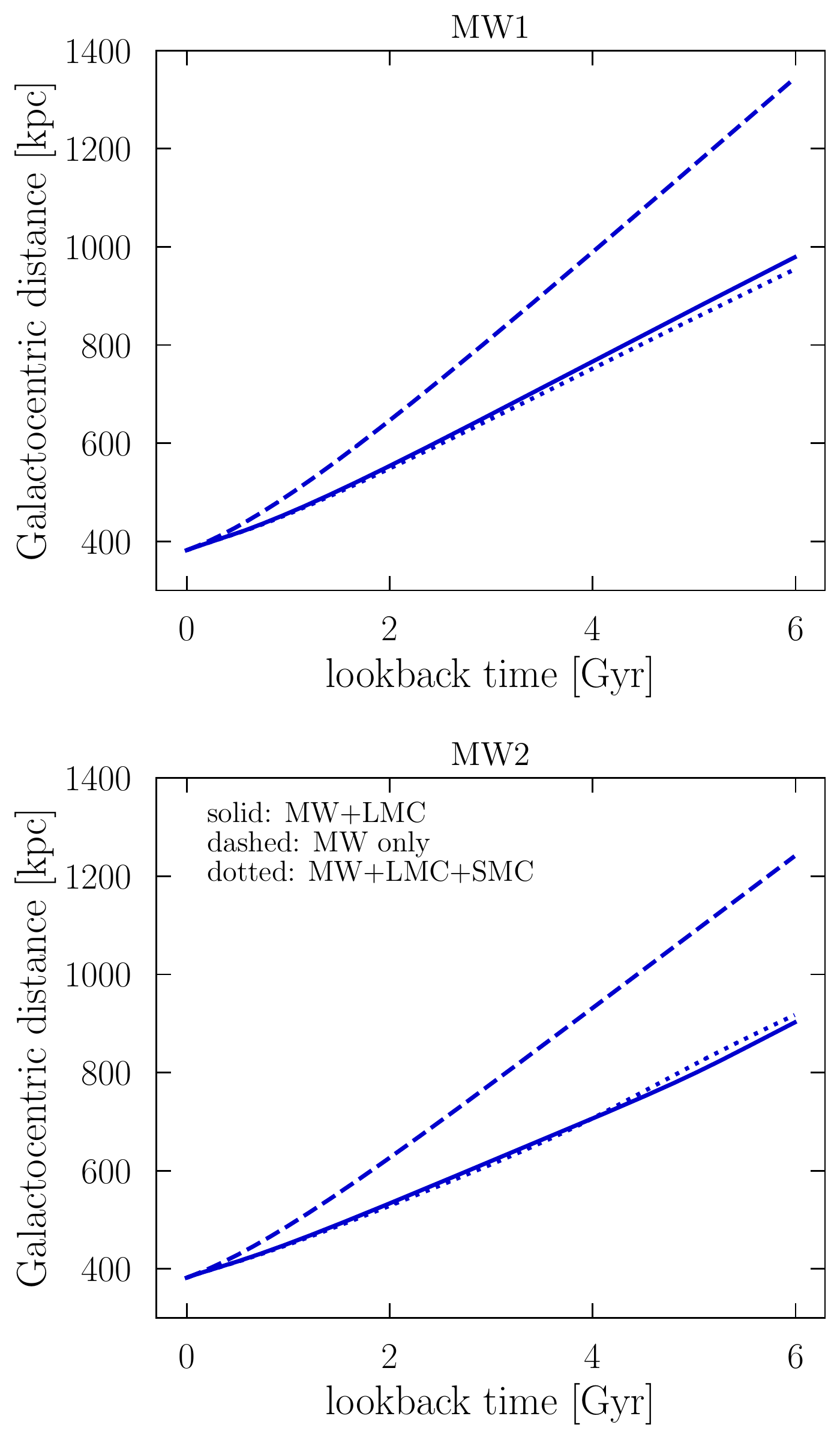}
    \caption{Orbital history of  of Eri~II in two different MW mass potentials using Gaia eDR3 proper motions \citep{mcconnachie21b}. MW1 has a virial mass of $10^{12}\, M_{\odot}$ while MW2 is more massive with a virial mass of $1.5 \times 10^{12}\, M_{\odot}$. Dashed lines represent orbits computed in a MW-only potential, solid lines indicate a MW+LMC potential with an LMC mass of $1.8\times 10^{11} \, M_{\odot}$, and dotted lines represent a MW+LMC+SMC potential with an SMC mass of $5\times 10^{9} \, M_{\odot}$. All of the above scenarios illustrate that Eri~II is on first infall into the Milky Way's halo.}
    \label{fig:EriIIorbit}
\end{figure}

\par In this section, we address the questions around quenching mechanisms for Eri~II. As part of this process, we derive the direct orbital history of Eri~II following the methodology outlined in \citet{patel2020}, using updated Gaia eDR3 proper motions from \citet{mcconnachie21b}, the radial velocity from \citet{li2017eriII}, and a distance consistent with that reported in Table \ref{tab:EriIIprop}. 

\par Figure \ref{fig:EriIIorbit} summarizes the orbital history of Eri~II in a low (MW1; $1\times10^{12}\,M_{\odot}$) and high mass (MW2; $1.5\times10^{12}\,M_{\odot}$) MW model. Orbits are computed in a MW-only (dashed lines), MW+LMC (solid lines), and a MW+LMC+SMC (dotted lines) potential. We only integrate our orbits as far back as 6~Gyr because attempts to ascertain the orbital history of Eri~II earlier than that will also need to account for the growth of the Milky Way potential, and because orbital uncertainties become significantly larger beyond 6 Gyr ago. 

\par In all scenarios, we find that Eri~II is on first infall into the halo of the MW. Following \citet{patel2020}, we also account for the measurement uncertainties in proper motion, line-of-sight velocity \citep{li2017eriII}, and distance by computing one thousand orbits using Monte Carlo drawings of the joint uncertainties as initial conditions. Regardless of assumed MW mass, our models suggest that Eri~II is statistically most likely to be on first infall as well (e.g., $\sim98$\% of all orbits result in the same orbital history). Moreover, we also verify that Eri~II does not pass within 800~kpc of M31 at any point in its trajectory. 

\par In summary, our updated orbit for Eri~II suggests that it was isolated throughout most of its evolutionary history, though we caution that this orbital history is still informed by large uncertainties on the proper motion of Eri~II. However, this result is also consistent with the results from \citet{battaglia2021}, who also derive an orbital history for Eri~II using Gaia eDR3 proper motions and find that it is on first infall into the MW. The revised orbital history of Eri~II from Gaia eDR3 proper motions is a departure from results of previous studies of Eri~II, which suggested that Eri~II has either completed multiple orbits around the MW (e.g., \citealt{li2017eriII}, \citealt{fritz2018}) or is a backsplash galaxy \citep{buck2019}. The discrepancies in these results are likely due to a combination of: 1. Different proper motion measurements between Gaia eDR3 and DR2 for Eri~II\footnote{As an example to illustrate this, the east component of Eri~II's proper motion changes between DR2 \citep{mcconnachie2020dr2} and DR3 \citep{mcconnachie21b} from $\mu_{\alpha}\mbox{cos}\delta=0.35^{+0.21}_{-0.20}$~mas~yr$^{-1}$ to $\mu_{\alpha}\mbox{cos}\delta=0.21\pm0.09$~mas~yr$^{-1}$ using the inference method that does not place prior expectations on the tangential velocity dispersion of the MW halo.}, 2. Orbit model-specific considerations (MW-only vs. including the Large Magellanic Cloud), and 3. Calculating the past orbit of Eri~II vs. orbital history inference by following Eri~II analogs in cosmological simulations.

\par We now turn to the question of how Eri~II quenched. At the time of its discovery, \citet{koposov2015DESsat} noted the presence of blue stars in the CMD of Eri~II that suggest Eri~II may have stopped forming stars only recently ($\sim250$~Myr ago). This scenario had also seemed viable because Eri~II is $\sim380$~kpc from the Milky Way, and thus less likely to have experienced ram pressure stripping that could have removed gas before the suspected burst of recent star formation. 

\par Subsequent studies, however, increasingly point away from this recent quenching scenario. \citet{rodriguezwimberly2019reionization} study Eri~II analogs in the Fat ELVIS simulations and find that it is unlikely for them to have quenched through infall into the Milky Way halo. The orbital history of Eri~II that we derive suggests that it fell into the Milky Way only recently, which does not correlate with deep HST studies of Eri~II suggesting that it formed the majority, if not all, of its stars before $z \sim 6$. 

\par Finally, if Eri~II was quenched through ram pressure stripping, we should expect to observe a sharp truncation in its MDF at the metal-rich end (e.g., the one-zone model of ram pressure stripping fit by \citealt{kirby2013LZR}), although accretion models with large $M$ could also produce a similar feature. While our MDF does not uniquely rule out ram pressure stripping, it adds to the growing body of evidence in the literature pointing away from ram pressure stripping as a possible source of quenching for Eri~II. 

\par Beyond that, the interpretation of Eri~II's old SFH is still up for debate. While \citet{simon2021EriII} interprets Eri~II's SFH as evidence of quenching by reionization, \citet{gallart2021eriii} suggests that SNe alone would be sufficient to quench Eri~II. The MDF of Eri~II that we derive does not provide definitive power to distinguish between these two mechanisms, as the one-zone models we fit do not distinguish between the details of how star-forming gas is depleted. However, any proper account of the physics driving star formation and quenching in Eri~II should also be able to reproduce its large internal metallicity spread. Similar theoretical benchmarks are also set by observations of the Segue~1 UFD, a much fainter UFD ($M_V=-1.3$) that appears to have experienced a short burst of star formation, but whose stars span almost 2 dex in metallicity \citep{frebel2014segue1}. 

\subsection{Environmental Impacts on Dwarf Galaxy Evolution}

\par In this section, we discuss the role of environment on dwarf galaxy evolution by comparing our results for Eri~II to two other galaxies that are close to Eri~II in luminosity, and which have published MDFs and one-zone chemical evolution model fits: Leo~T, and Boo~I. 

\par  Leo~T is similar to Eri~II in luminosity ($L\sim~4\times10^4~L_{\odot}$; \citealt{irwin2007}) and distance ($d_{MW}\sim410$~kpc; \citealt{clementini2012}) from a more massive host. The key distinction between these two galaxies is that Leo~T displays an extended SFH that continues past reionization (\citealt{clementini2012}, \citealt{weisz2012}), with a substantial reservoir of H~I gas ($M\sim2.8\times10^5~ M_{\odot}$; \citealt{ryan-weber2008}), while Eri~II stopped forming stars 13~Gyr ago. The orbital history of Leo~T suggests that like Eri~II, it also seemed to have evolved in isolation \citep{battaglia2021}. Finally, similar to Eri~II, the MDF of Leo~T is best described by a Leaky Box \citep{kirby2013LZR}. 

\par On the other hand, Boo~I has a CMD which suggests that it underwent a similar SFH to Eri~II: a short starburst that ended also around 13~Gyr ago \citep{brown2014sfh}. Unlike Eri~II, the orbital history of Boo~I suggests that it is bound to the Milky Way (\citealt{simon2018}, \citealt{fritz2018}) and, as a result, possibly formed in a denser environment relative to that of Eri~II.\footnote{Although, \citet{fillingham2019} show that it is also possible for the star formation epoch of Boo~I to have ended prior to infall into the Milky Way.} Finally, the MDF of Boo~I is best represented by an Accretion model \citep{jenkins2021vlt}, suggesting that the dense environment it was born in could have allowed it access to extra gas.

\par As a point of reference, \citet{kirby2013LZR} fit one-zone models to the MDFs of the more massive Local Group dwarf galaxies, and found that while the MDFs of dwarf spheroidal galaxies, which are bound to the Milky Way, are better described by Accretion models, the MDFs of isolated dIrrs are better described by Leaky Box or Pre-Enriched models. To the best of the data that we currently have for galaxies of lower luminosities, these distinctions between MDFs seems to also hold for UFDs that have long been bound to a more massive host versus UFDs that likely evolved in more isolated environments. 

\par Instead of present-day morphology, \citet{gallart2015} offer a framework for interpreting the role of environment in dwarf galaxy formation by classifying dwarf galaxies into \textit{fast dwarfs} and \textit{slow dwarfs} based on their SFH. \textit{Fast dwarfs} are those with SFHs consisting of a single short, but dominant, event. \textit{Slow dwarfs} show current or recent SFH, with continuous activity starting from the oldest epochs. Under this framework, distinctions in SFH arise from the environments in which the galaxies were forming stars: \textit{fast dwarfs} formed in dense environments where they frequently experienced interactions that would result in rapid star formation (e.g., gas accretion, mergers) and its rapid truncation (e.g. feedback from SNe and reionization from within and from adjacent galaxies), while \textit{slow dwarfs} formed throughout cosmic time in a lower-density environment far from the influence of massive hosts. 

\par While the characteristics of Boo~I and Leo~T respectively fit into the \textit{fast dwarfs} and \textit{slow dwarfs} categories, Eri~II complicates this picture: it resembles a \textit{fast dwarf} on the basis of its short and early burst of SFH, but the updated orbit that we derive for Eri~II and the similarity of its MDF to those of several other isolated dIrrs suggests that Eri~II has largely evolved in isolated like \textit{slow dwarfs}. 

\par Using the SFH of Eri~II, \citet{gallart2021eriii} argue that it is possible for Eri~II to self-quench entirely through SNe feedback rather than reionizing radiation, e.g. from adjacent massive hosts. If this is the case, then UFDs with SFHs that would render them as \textit{fast dwarfs} can also quench early, independent of environment. On the other hand, the similarity between the MDFs of Eri~II and Leo~T, and of some other isolated dIrrs, despite their different SFHs, raises the possibility of common formation pathways for \textit{fast} and \textit{slow} dwarfs that evolve in isolation.

\par We of course offer the above discussions with the caveat that they consider limited numbers of UFDs, and with results that would benefit from further refinement; e.g. the Leaky Box fit for Leo~T was made using only 16 stars. In any case, this line of investigation could be furthered with studies of more UFDs. Here, the CaHK imaging technique will also be useful, as it will allow us access to stellar metallicities of the less luminous and more distant galaxies currently known (e.g. Andromeda~XVI, as part of HST-GO-16686; PI D. Weisz), and that are expected to be discovered in upcoming surveys. 

\subsection{Interpretation of UFD MDFs in Cosmological Simulations}

\par While one-zone chemical evolution models have been pivotal in building intuition around the relationship between galaxy MDFs and the physics driving their evolution, detailed simulations show that a subset of dwarf galaxies are affected by a combination of reionization (e.g., \citealt{dawoodbhoy2018reionization}, \citealt{graus2019reionization}), local environment (e.g., \citealt{applebaum2021}), and internal stellar feedback (\citealt{sun2018discreteSNe}, \citealt{smith2019feedback}). While some studies have produced certain observed features, such as the SFH of isolated dwarf galaxies (e.g., \citealt{fitts2017fieldgalaxies}, \citealt{revaz2018}), and have begun to resolve bulk properties such as mean metallicity in UFDs \citep[e.g., ][]{wheeler2019ufdsims}, the lack of robust MDF observations of UFDs have motivated few studies to extensively simulate UFD MDFs as tracers of their formation processes. \citet{jeon2017popIII} is one of the few studies so far that predict MDFs of UFDs by following the evolution of 6 halos with present-day stellar masses from $10^4~M_{\odot}$ to almost $10^6~M_{\odot}$. We use their MDFs as a point of comparison.

\par
\begin{figure*}
    \centering
    \includegraphics[scale=0.35]{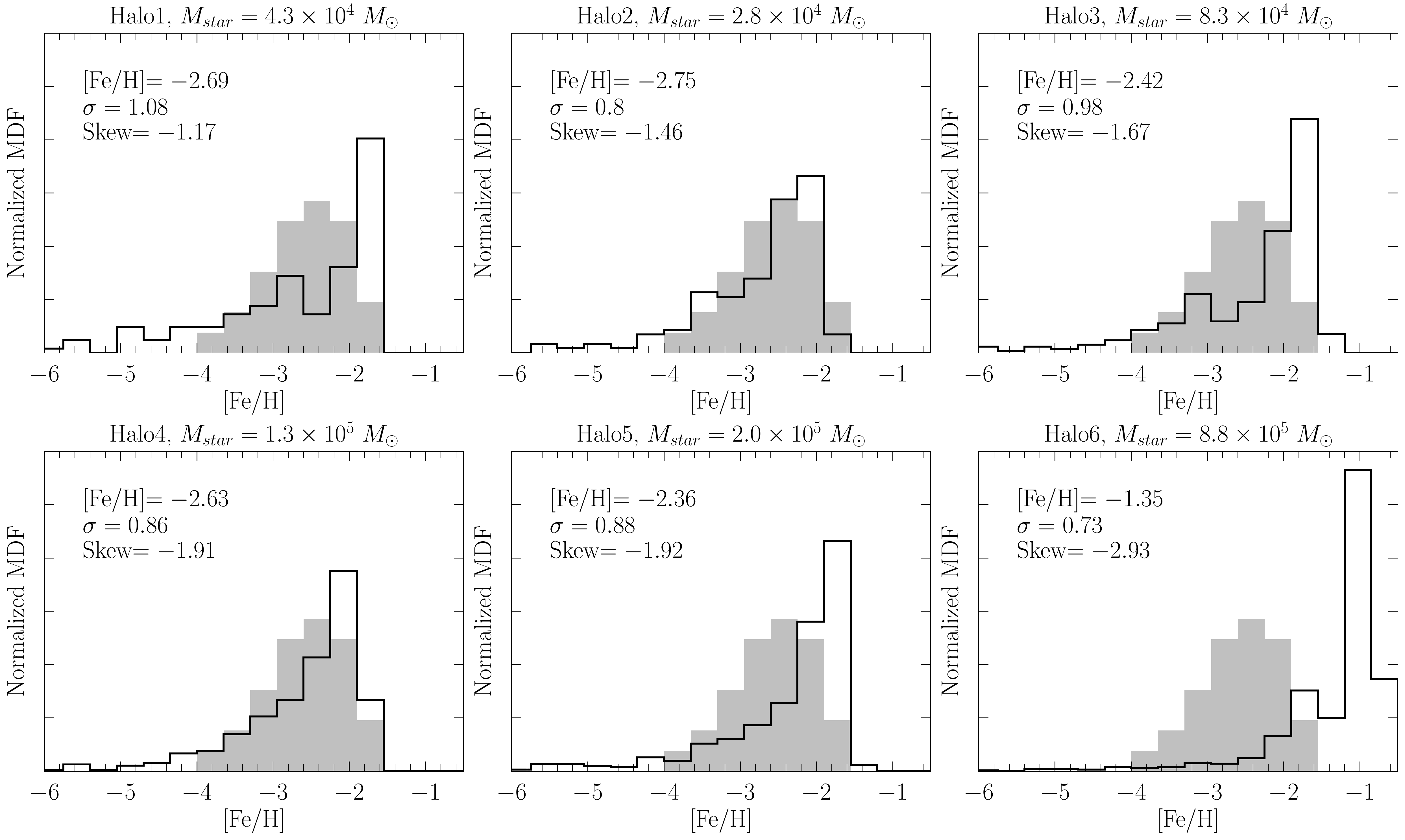}
    \caption{Comparison of the MDF that we derive for Eri~II (gray shaded histogram; $M_{star}\sim1\times10^5M_{\odot}$, $\mbox{[Fe/H]}=$\met, $\sigma=\metdisp$, $\mbox{skew}=-0.21$) with the 6 UFD MDFs simulated from \citet{jeon2017popIII}. Following convention in previous histogram plots, binsizes are 0.35~dex wide. The stellar mass of the corresponding halo at $z=0$ are in the title of each panel. We also compute the mean metallicity, metallicity dispersion, and skew of the simulated MDFs. All of the simulated MDFs are skewed relative to Eri~II, with long metal-poor tails.}
    \label{fig:jeon17comparison}
\end{figure*}

 Figure \ref{fig:jeon17comparison} presents the MDF of Eri~II, compared to the MDFs of the 6 simulated UFDs from \citet{jeon2017popIII}. Halos labeled Halo1, Halo2, and Halo3 are UFD analogs whose star formation was ultimately shut off after reionization, though the study finds that both SNe and reionization are necessary to quench these galaxies. We consider these halos to have a SFH that is most analogous to that of Eri~II. The remaining massive halos simulated were able to form stars after reionization through gas accretion and mergers. Comparing the MDF of the first 3 halos to Eri~II, Halo2 has the MDF that is most similar to that of Eri~II, although within the simulation there is still variation in MDFs between halos that experienced similar SFHs and quenching mechanisms. 

\par Regardless of mass, all of the simulated MDFs display long, metal-poor ($\mbox{[Fe/H]}<-4.0$) tails. In the simulation, these stars represent Pop~II stars that were directly enriched by the death of Pop~III stars at $z>11$ in mini halos that later merged to form the final UFD analogs. This interpretation of the origin of the metal-poor tail is in contrast with the results of our one-zone fit to Eri~II's MDF, which also predicts a long, metal-poor tail, but which attributes it through chemical enrichment taking place in a single system rather than in the aggregate of many. Additional theoretical work to refine the interpretation of various components of MDFs would be of great interest in understanding small-scale galaxy formation. In any case, both comparison to one-zone models and cosmological simulations point to additional theoretical basis for expecting to find EMP stars in Eri~II, and possibly in other UFDs as well. CaHK imaging studies with larger FoVs and increasing depths could uncover additional EMP candidates, but since these stars would have metallicities at the limit of discernment via CaHK, future discoveries would need to confirmed via low-resolution spectroscopic studies on current facilities or observations on the next generation of ELTs \citep[see, e.g., ][]{sandford2020}.

\par Finally, for each of the simulated MDFs, we also compute their mean metallicity, metallicity dispersion, and skew. The skew metrics confirm what we also verify visually: that all of the simulated MDFs from \citet{jeon2017popIII} are left-skewed compared to that of Eri~II, in part because our models do not reach below $\mbox{[Fe/H]}=-4.0$. As a result, the simulated MDFs tend to have a lower average metallicity than Eri~II, and also larger metallicity dispersions. 

\par Since there are stars in our sample that may be more metal-poor than $\mbox{[Fe/H]}=-4.0$ to within 2 sigma, improved metallicity measurements may populate the metal-poor tail of Eri~II's MDF. If Eri~II also has a radial metallicity gradient decreasing outwards, then measuring the metallicity of Eri~II members out to larger radii may also bring the MDF of Eri~II closer to predictions from this study. 

\par More generally, since a viable theory of galaxy formation should be able to produce features of the UFD population, we suggest that future population-level comparisons between simulated and observed UFD MDFs could also account for MDF statistics. While the population of mean metallicity and metallicity dispersions would be useful baseline statistics to compare, we also suggest that higher-order moments such as skew and kurtosis would be important for capturing asymmetries in the shape of UFD MDFs, which are currently also represented in the one-zone models. 

\section{Conclusion}
\label{sec:conclusion}

\par In this work, we combined HST WFC3/UVIS imaging in the narrow-band $F395N$ filter with broadband archival photometry to measure metallicities for 60 stars in Eri~II, increasing the sample of member stars with known metallicities by a factor of nearly 4. We derive a mean metallicity of $\mbox{[Fe/H]}=$\met$^{+\metuerr}_{-\metlerr}$ and metallicity dispersion $\sigma_{\mbox{[Fe/H]}}=\metdisp^{+\metdispuerr}_{-\metdisplerr}$, which is consistent with results from spectroscopic studies and with known trends about UFDs. We are able to independently confirm that Eri~II is a UFD whose mean metallicity is consistent with properties of the general UFD population, and that it has a large internal metallicity dispersion. These results affirm the promise for HST Ca H\&K narrow-band photometry to faithfully recover the MDF properties of Eri~II's fainter UFD counterparts.

\par From an expanded sample of stellar metallicities, we fit one-zone chemical evolution models to the MDF of Eri~II and find that it is best-represented by a Leaky Box, similar to results from studies of LG isolated dwarf galaxies. We also present an updated orbit for Eri~II, which suggests that it only recently fell into the Milky Way. The composite of these results suggest that Eri~II likely formed in an underdense region of the universe at high redshifts. 

\par We also use our photometry to identify outliers in our derived MDF that may be promising targets for follow-up spectroscopy to 1) verify kinematic membership with Eri~II, 2) refine the shape of the Eri~II MDF at the most metal-poor end, and 3) conduct detailed chemical abundance analysis to infer sources of past enrichment events. Since the field of our narrow-band imaging also includes the central cluster of Eri~II, future work will also involve measuring the metallicity of the cluster as well. 

\par  Finally, having verified the efficacy of Ca H\&K narrow-band imaging for inferring the MDF of UFDs, our forthcoming publications will characterize the MDFs of the rest of the 18 UFD candidates observed by HST-GO-15901. 

\acknowledgements

\par We thank Aaron Dotter for providing advance access to the MIST v2 models that include $\alpha$ enhancement. We thank L\'eo Girardi for adding $F395N$ to the available set of simulated filters in the TRILEGAL models for our work. We thank Clara Mart{\'\i}nez-V{\'a}zquez for providing metallicity data from \citet{martinezvazquez2021} for evaluation against our measurements. We thank Myoungwon Jeon for making the MDFs from \citet{jeon2017popIII} available for our analysis. We also thank Ani Chiti, Andrew Cole, Ting Li, Josh Simon, Evan Skillman, and Kim Venn for useful conversations about this work. SWF, DRW, MBK, APJ, and NRS are thankful for the stimulating discussion at The Intersection of Age Measurements Across the Universe workshop hosted in Napa, CA in December 2019. Finally, we thank the anonymous referee for comments that improved the quality of this manuscript.

\par SWF acknowledges support from a Paul \& Daisy Soros Fellowship, as well as from the NSFGRFP under grants DGE 1752814 and DGE 2146752. SWF also acknowledges support from HST-GO-15901, HST-GO-16226 from the Space Telescope Science Institute, which is operated by AURA, Inc., under NASA contract NAS5-26555. DRW acknowledges support from HST-GO-15476, HST-GO-15901, HST-GO-15902, HST-AR-16159, and HST-GO-16226 from the Space Telescope Science Institute, which is operated by AURA, Inc., under NASA contract NAS5-26555. ES acknowledges funding through VIDI grant ``Pushing Galactic Archaeology to its limits" (with project number VI.Vidi.193.093) which is funded by the Dutch Research Council (NWO). NFM gratefully acknowledges support from the French National Research Agency (ANR) funded project ``Pristine'' (ANR-18-CE31-0017) and from the European Research Council (ERC) under the European Unions Horizon 2020 research and innovation programme (grant agreement No. 834148). MBK acknowledges support from NSF CAREER award AST-1752913, NSF grant AST-1910346, NASA grant NNX17AG29G, and HST-AR-15006, HST-AR-15809, HST-GO-15658, HST-GO-15901, HST-GO-15902, HST-AR-16159, and HST-GO-16226 from the Space Telescope Science Institute, which is operated by AURA, Inc., under NASA contract NAS5-26555. 

\software{DOLPHOT (\citealt{dolphot2016}, \citealt{hstphot2000}), astropy  \citep{astropy}, numpy \citep{numpy}, matplotlib \citep{matplotlib}, emcee \citep{emcee}, corner \citep{corner}, scipy \citep{scipy}, dynesty \citep{dynesty}}

\bibliography{bibliography}{}
\bibliographystyle{aasjournal}

\appendix

\section{Impact of alpha element abundances on CaHK photometry}
\label{sec:appendix_alphaeffect}

\par
\begin{figure*}
    \centering
    \includegraphics[scale=0.38]{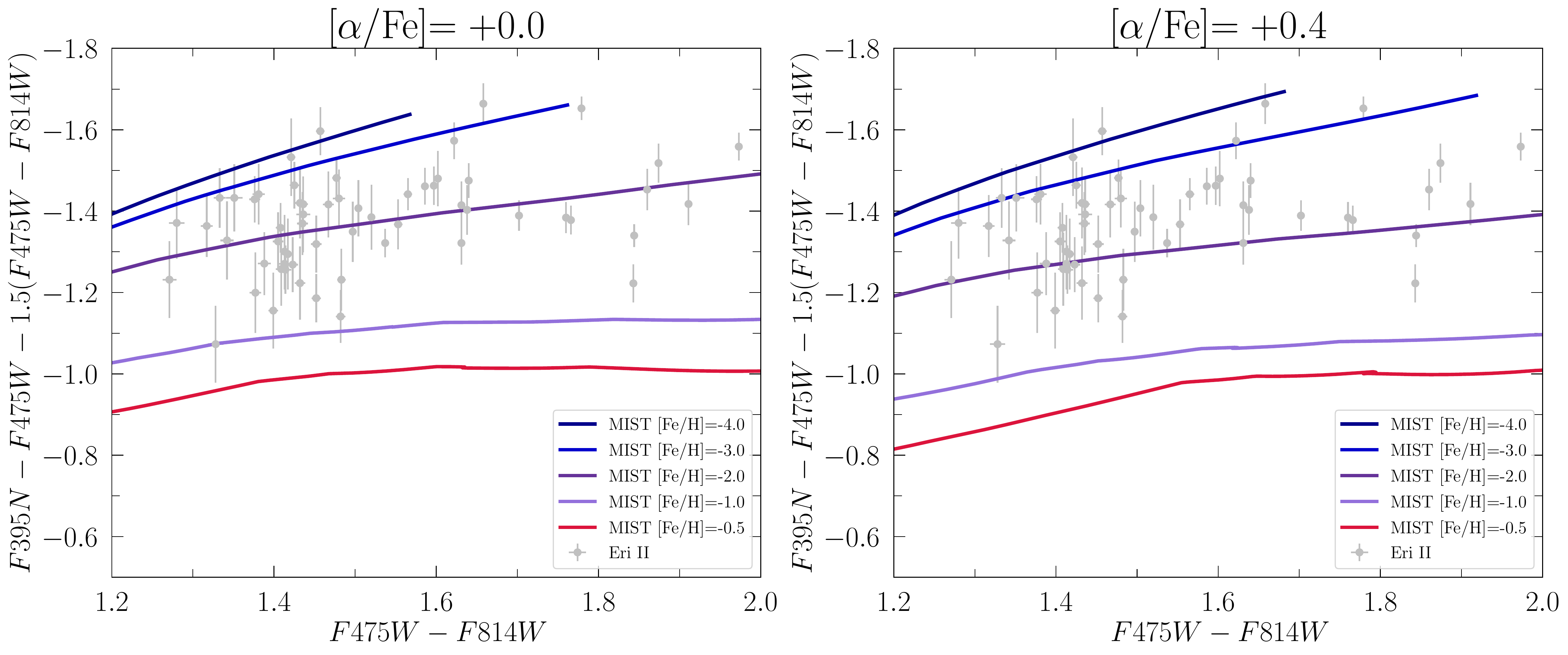}
    \caption{Comparing MIST isochrones in CaHK color space for isochrone models with [$\alpha$/Fe]$=0.0$ (left panel) vs [$\alpha$/Fe]$=+0.40$ (right panel). For an isochrone of a given metallicity, the $\alpha$-enhanced model reaches redder extents in $F475W-F814W$, and is also redder in CaHK color index.}
    \label{fig:mist_alpha}
\end{figure*}

 Figure \ref{fig:mist_alpha} compares the MIST models in CaHK color space for Solar-scaled alpha abundances (left) to those for [$\alpha\mbox{/Fe]}=+0.4$ (right). At a given metallicity, the alpha-enhanced tracks are shifted down (redward) in CaHK space relative to the Solar-scaled alpha tracks. For the more metal-poor models, the alpha-enhanced models also extend redder in $F475W-F814W$. Stars which fall outside the Solar-scaled alpha grid are within the alpha-enhanced models. 

\par If we re-compute our MDF using the Solar-scaled alpha models, we derive a mean metallicity of $\mbox{[Fe/H]}=-2.26\pm0.07$, and a metallicity dispersion of $\sigma_{\mbox{[Fe/H]}}=0.44\pm0.06$. Thus, the MDF we derive would be more metal-rich, but the metallicity dispersion would remain unchanged. While alpha abundances changes our inferred stellar metallicities, these differences are not enough to account for the discrepancies between our measurements and those of Z20, since as Figure \ref{fig:indiv_histograms} shows, the discrepancies in measurements are at least 1~dex or more. 

\section{Select Metallicity Posterior Distributions of Eri~II Stars}
\label{sec:appendix_posteriordistrs}

\par In this section, we present select marginalized posterior distributions that result from our measurements of individual stellar metallicities. 

\par
\begin{figure}
    \centering
    \includegraphics[scale=0.4]{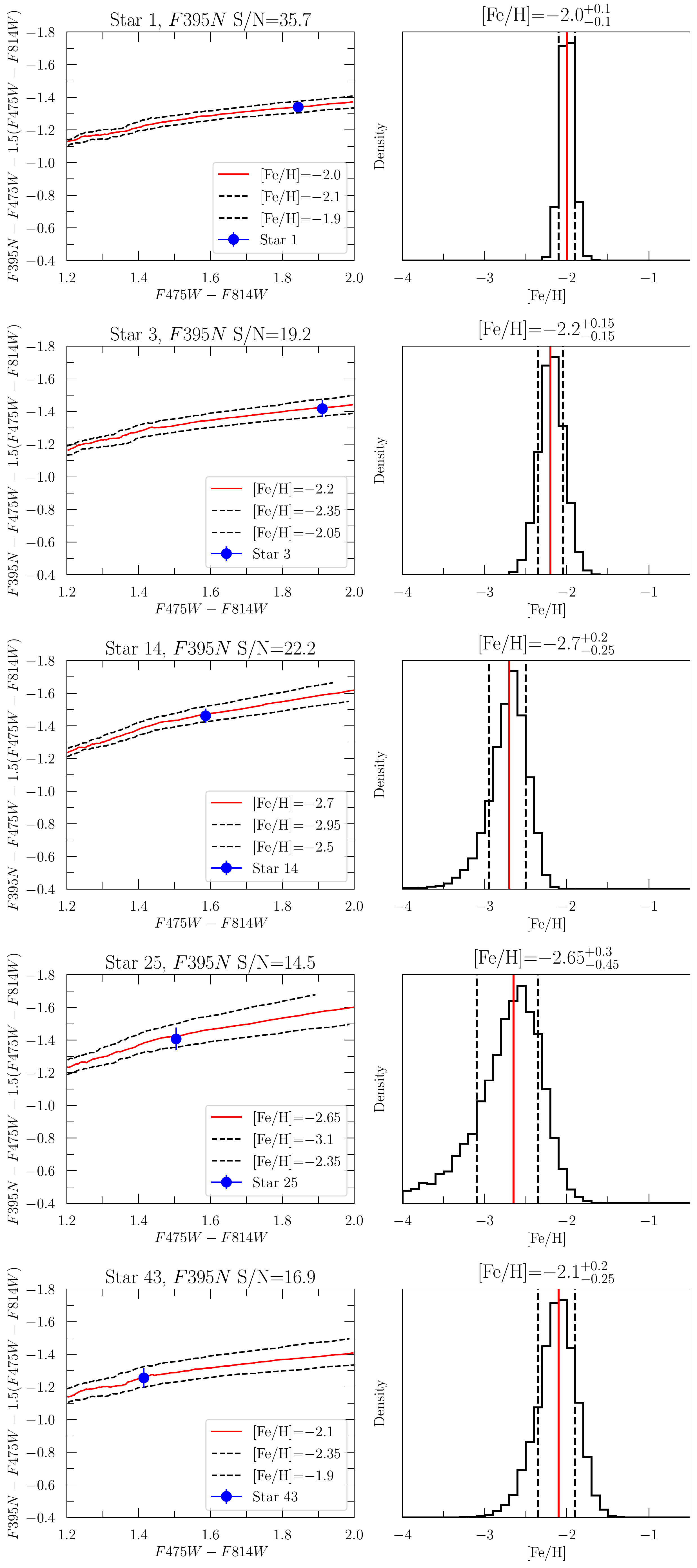}
    \caption{Well-constrained posterior distributions.}
    \label{fig:normal_posteriors}
\end{figure}

 Figure \ref{fig:normal_posteriors} presents well-constrained marginalized posterior distributions. These distributions have a clear peak with clear tails. Because CaHK is less effective at distinguishing between lower metallicities (e.g., Figure \ref{fig:mist_alpha}), the tails of these posterior distributions are often longer at the metal-poor end. 

\par
\begin{figure}
    \centering
    \includegraphics[scale=0.4]{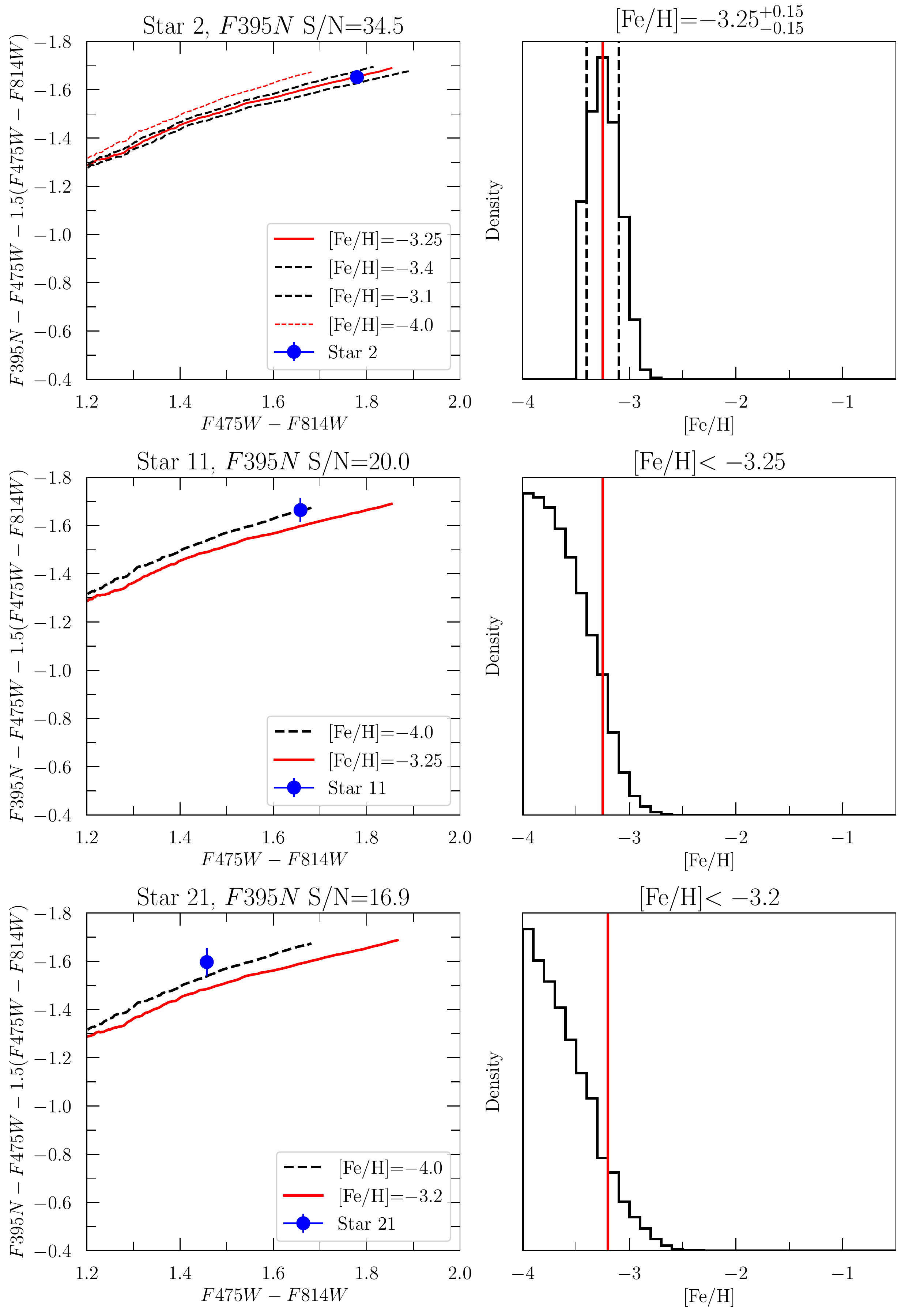}
    \caption{Posterior distributions of stars that we flag as spectroscopically accessible EMP candidates for follow-up studies, and which we discuss in Section \ref{sec:noteworthystars}. }
    \label{fig:mp_posteriors}
\end{figure}

 Figure \ref{fig:mp_posteriors} shows the posterior distributions of stars that we identify as promising candidates for spectroscopic follow-up in Section \ref{sec:noteworthystars} and Table \ref{tab:indiv_measurements}. 

\par
\begin{figure}
    \centering
    \includegraphics[scale=0.4]{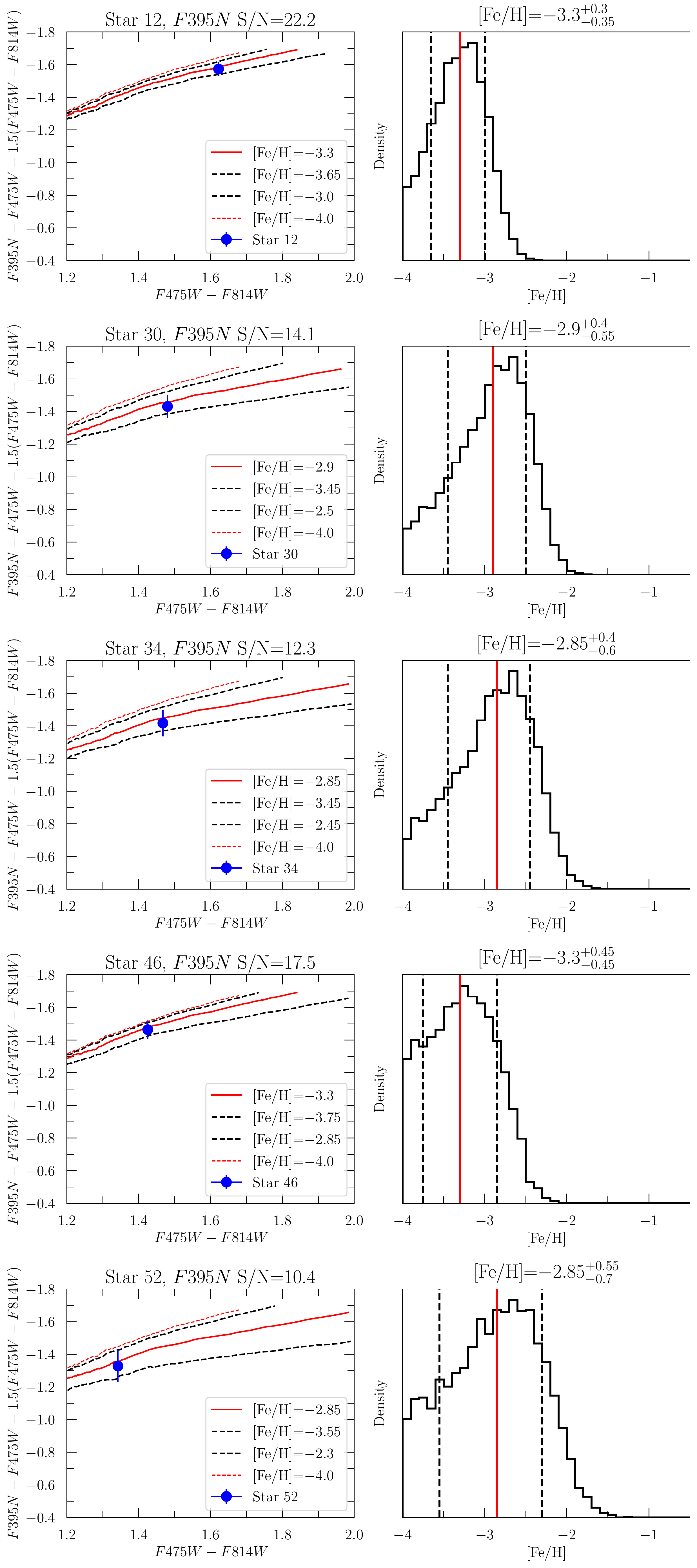}
    \caption{Posterior distributions of stars that are truncated at the 2 sigma level on the metal-poor end. The thin dotted red line is the most metal-poor ($\mbox{[Fe/H]}=-4.0$ model in our grid. }
    \label{fig:truncated_posteriors}
\end{figure}

 Figure \ref{fig:truncated_posteriors} shows the posterior distributions of stars whose 2 sigma metal-poor tail extends beyond the lowest metallicity of our grid. 

\section{Recovery of Synthetic MDF}

\par
\begin{figure*}
\includegraphics[scale=0.45]{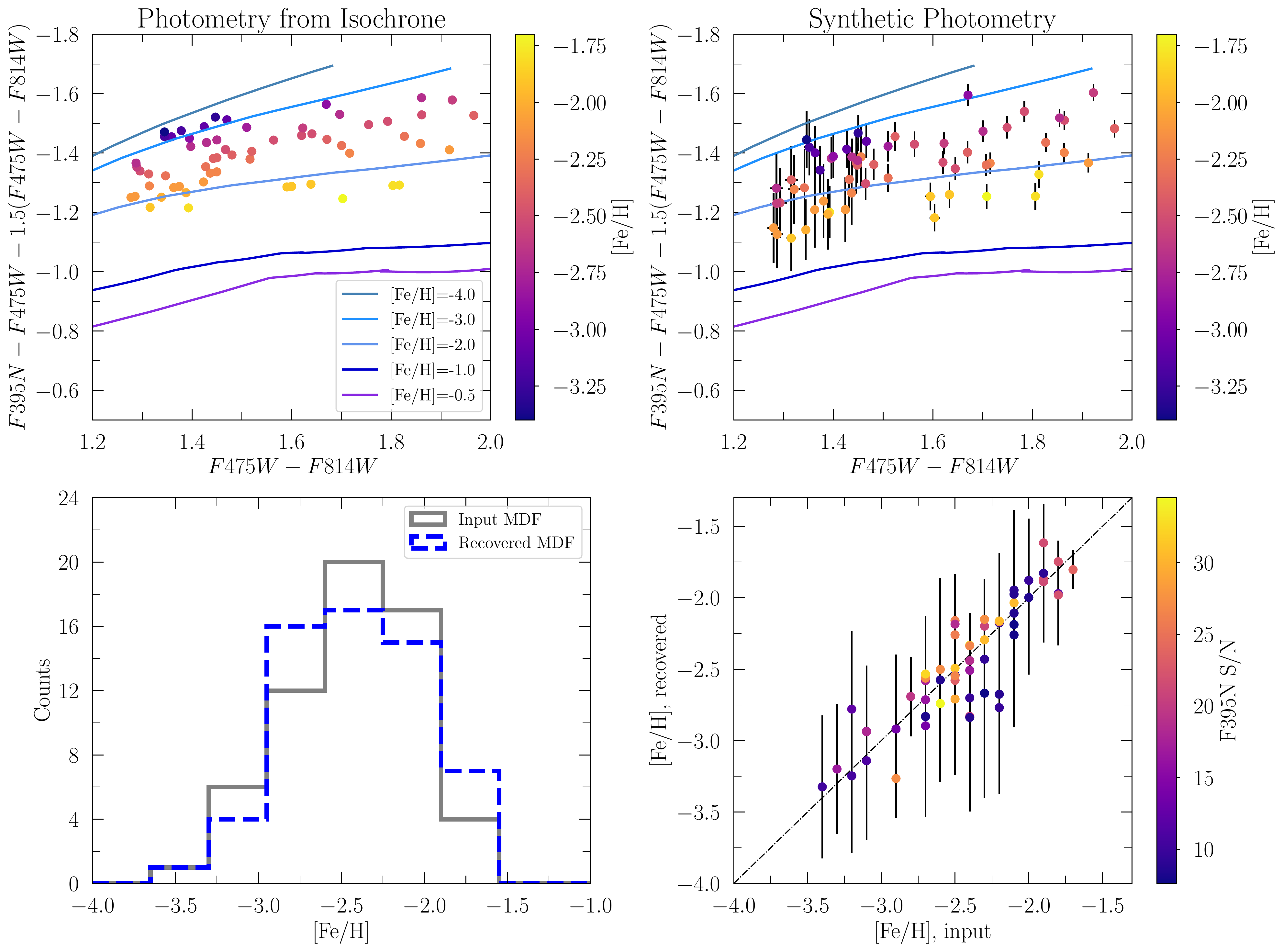}
\caption{\textbf{(top left)} The 60 model photometry points that we draw from isochrone models, with metallicity counts set by a Gaussian MDF with a mean and spread consistent with what we measure for Eri~II. Points are color-coded by their known metallicities. We also apply extinction to the tracks and the photometry in this plot, but it is negligible (on order of $\sim$0.01~mag). \textbf{(top right)} Synthetic photometry after appying bias and uncertainty from the ASTs (described in Section \ref{sec:data} to the model photometry. As in the previous panel, points are color-coded by their known metallicities. \textbf{(bottom left)} Comparing the input MDF to the MDF recovered after measuring individual stellar metallicities of the synthetic photometry, following the procedures in Section \ref{sec:metallicity_indiv}. \textbf{(bottom right)} Comparing the input metallicity to the recovered metallicity measurements for individual stars. Points are color-coded by their synthetic F395N S/N, as determined from the ASTs. The dash-dotted line is a 1-1 relation.}
\label{fig:syntheticmd}
\end{figure*}

 In this section, we demonstrate the efficacy of the CaHK stellar metallicity inference methods applied in Section \ref{sec:metallicity_indiv} by recovering a theoretical MDF from synthetic photometry. Figure \ref{fig:syntheticmd} summarizes the results of this exercise.

\par First, we draw 60 stars from a Gaussian MDF with a mean and spread set by the values that we infer for Eri~II in Section \ref{sec:resultsgauss}. We use the metallicities of those stars to draw model photometry from isochrones of corresponding metallicity, and ensure that the synthetic colors fall within the CaHK color space range set by our Eri~II data. The top left panel of Figure \ref{fig:syntheticmd} shows the location of the resulting model photometry in CaHK color space.

\par To arrive at the final catalog of synthetic photometries, we apply a bias and error to each model photometry point following its corresponding error profile as determined by the ASTs from Section \ref{sec:data}. We apply bias as an offset, and apply error as a random draw from a Gaussian corresponding to its AST uncertainty. We also adopt the error bar from our ASTs as the corresponding uncertainties for each synthetic photometric point. The top right panel of Figure \ref{fig:syntheticmd} shows the final synthetic photometry in CaHK color space, with their corresponding uncertainties. 

\par We then measure metallicites from the synthetic photometry following the procedures of Section \ref{sec:metallicity_indiv}. The bottom left panel of Figure \ref{fig:syntheticmd} compares the input MDF to the recovered MDF, and the bottom right panel of Figure \ref{fig:syntheticmd} compares the input metallicities to the recovered metallicities for individual stars. We verify that we can recover the input MDF and individual stellar metallicities following our method for inferring the MDF of Eri~II.

\section{Posterior Distributions of Analytic MDF Parameters}

\begin{figure}
    \centering
    \includegraphics[scale=0.5]{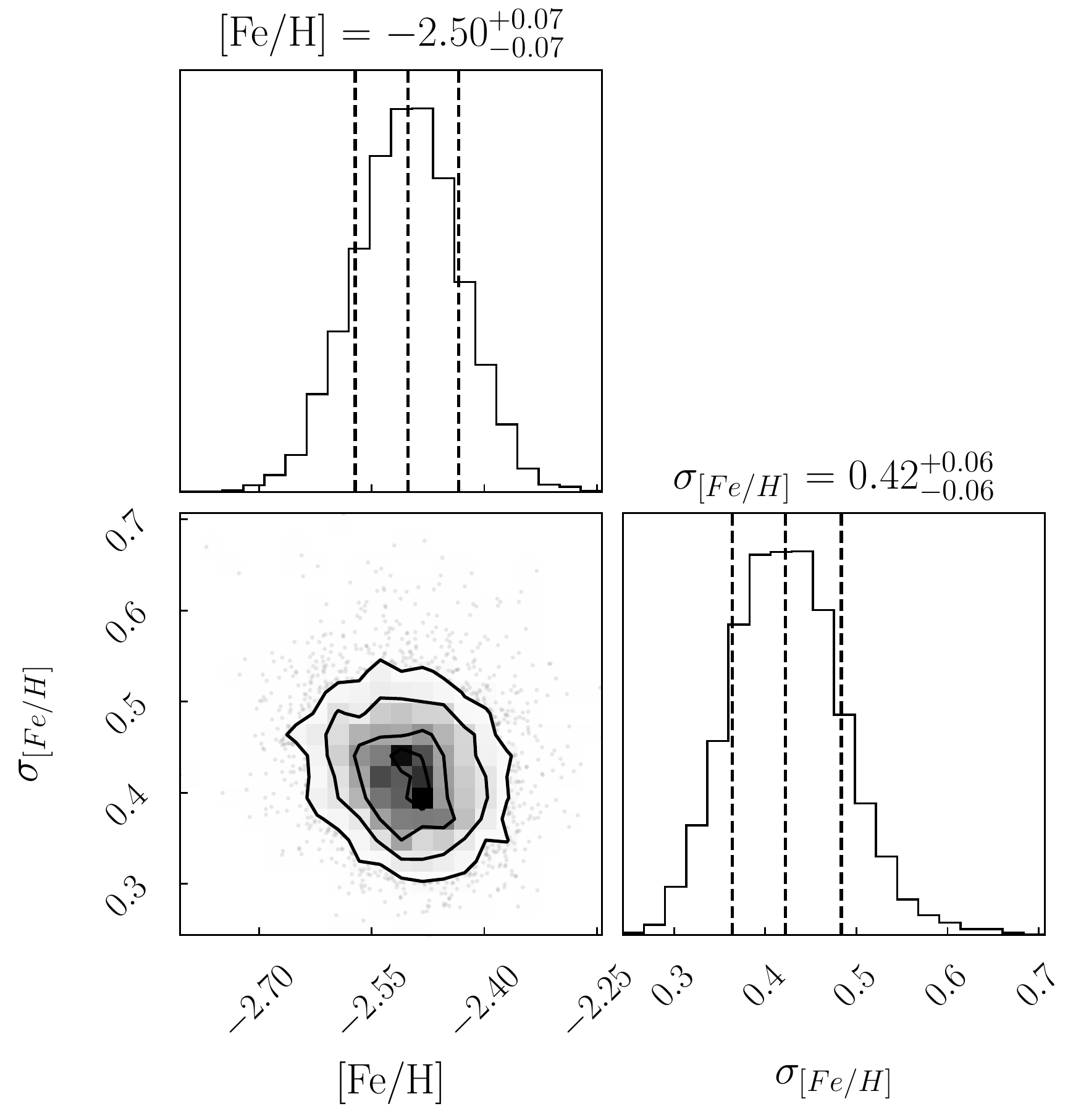}
    \caption{Joint distribution of the mean metallicity ([Fe/H]) and metallicity dispersion ($\sigma_{\mbox{[Fe/H]}}$) of Eri~II. The dashed lines mark the 16th, 50th, and 84th percentile of the 1D posterior distributions.}
    \label{fig:gaussian_corner}
\end{figure}

\begin{figure}
    \centering
    \includegraphics[scale=0.55]{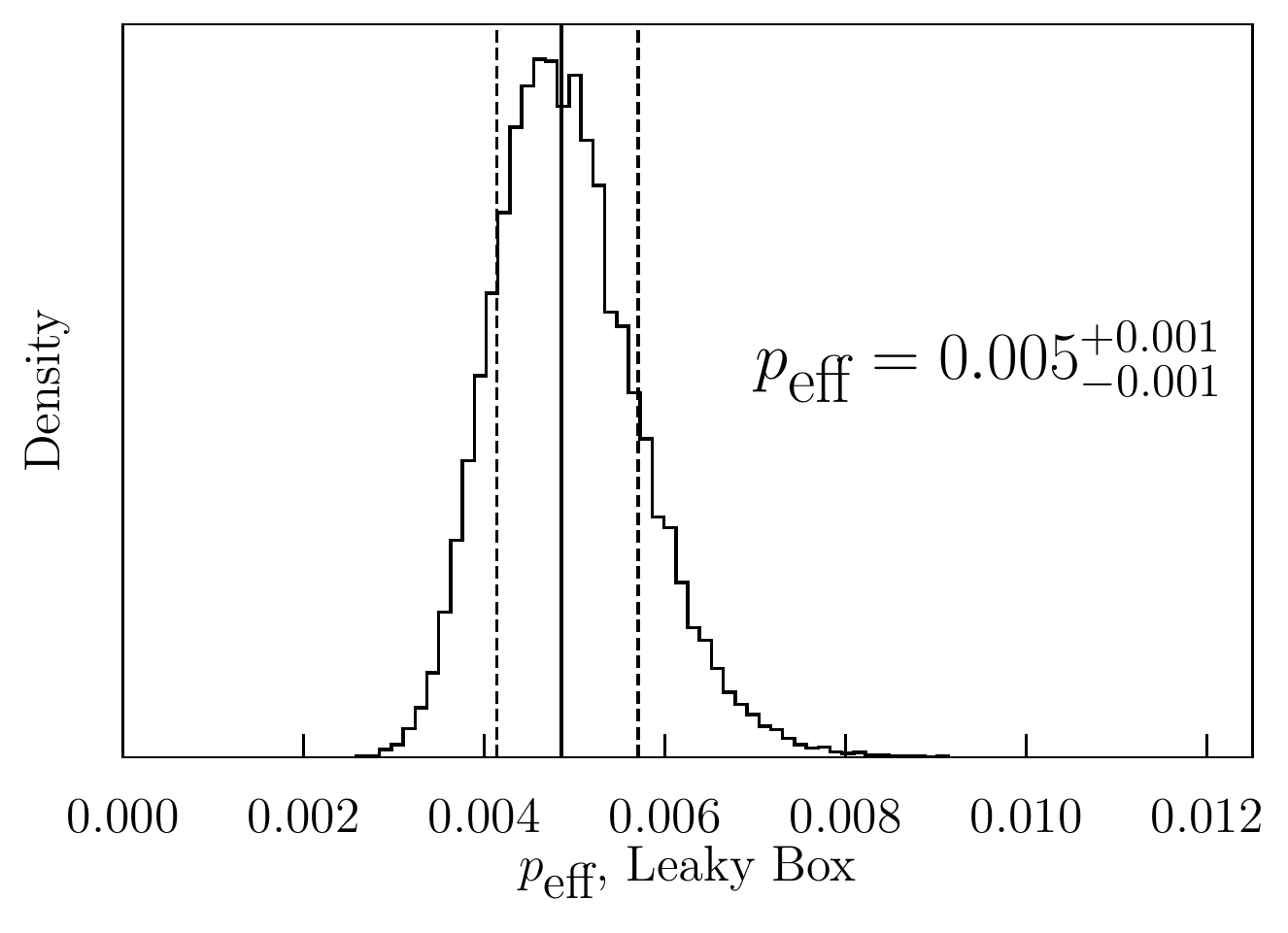}
    \caption{Posterior distribution of $p_{\mbox{eff}}$ for the Leaky Box model.}
    \label{fig:leakybox_posterior}
\end{figure}

\begin{figure}
    \centering
    \includegraphics[scale=0.53]{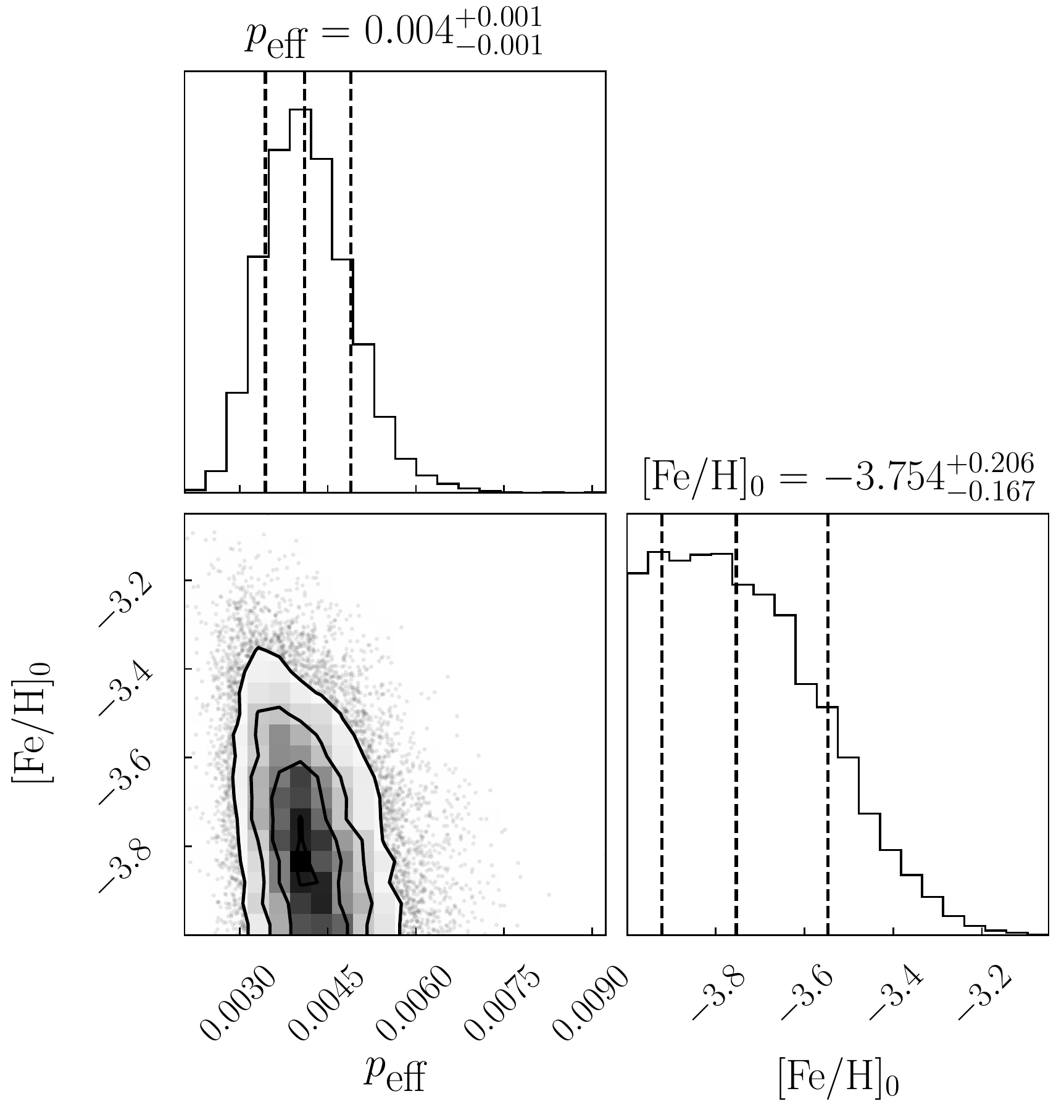}
    \caption{Posterior distribution corner plots of $p_{\mbox{eff}}$ and [Fe/H]$_{0}$ for the Pre-Enriched gas model.}
    \label{fig:preenriched_posterior}
\end{figure}

\begin{figure}
    \centering
    \includegraphics[scale=0.53]{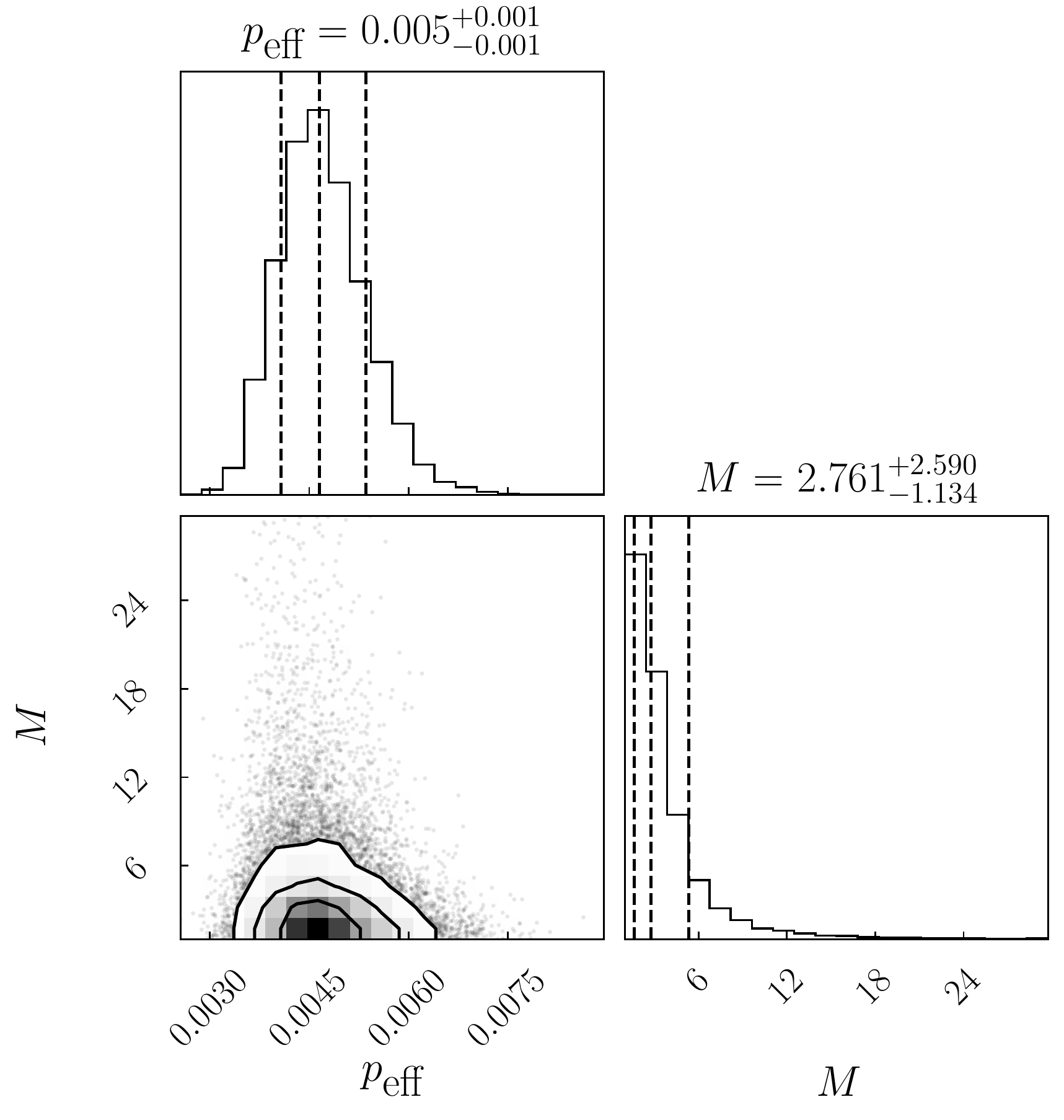}
    \caption{Posterior distribution corner plots of $p_{\mbox{eff}}$ and $M$ for the Accretion model.}
    \label{fig:extragas_posterior}
\end{figure}

\section{Table of measurements}

\begin{longrotatetable}
\begin{deluxetable*}{cCCCCCCCCCc}
\tablecaption{Metallicity Measurements for Eri~II stars. The columns are as follows: (1) Internal Star ID; (2) \& (3) R.A. and Decl.; (4) F475W mag; (5) F475W-F814W color; (6) CaHK index defined in Section \ref{sec:data}; (7) [Fe/H] measurement based off the 16th, 50th, and 84th percentiles of each star's PDF; (8) Upper limit on [Fe/H] where relevant; (9) metallicity measurement from L17 where available; (10) metallicity measurement from Z20 where available}
\label{tab:indiv_measurements}
\tabletypesize{\scriptsize}
\tablehead{
\colhead{StarID}    & \colhead{R.A. (J2000)} & \colhead{Decl. (J2000)} & \colhead{$F475W$} & \colhead{$F475W - F814W$} & \colhead{$CaHK$} &  \colhead{[Fe/H]$_{perc}$} & \colhead{[Fe/H]$_{ulim}$} & \colhead{[Fe/H], L17}    &  \colhead{[Fe/H], Z20} \\ 
\colhead{}         & \colhead{(deg)} & \colhead{(deg)} &  \colhead{(mag)} &   \colhead{(mag)} &   \colhead{(mag)} &            \colhead{(dex)} &       \colhead{(dex)} & \colhead{(dex)}          &  \colhead{(dex)}       } 
\startdata
                  0 &       56.09995 &       -43.54543 &          21.243 &    1.973 \pm 0.003 & -1.558 \pm  0.034 &      -2.55^{+0.10}_{-0.11} &                   ... &      -2.18 \pm      0.14 &                    ... \\ 
                  1 &       56.11482 &       -43.52685 &          21.251 &    1.844 \pm 0.002 & -1.340 \pm  0.028 &      -2.00^{+0.08}_{-0.08} &                   ... &                      ... &                    ... \\ 
 2\tablenotemark{a} &       56.12385 &       -43.52998 &          21.288 &    1.779 \pm 0.002 & -1.653 \pm  0.029 &      -3.24^{+0.16}_{-0.15} &                   ... &      -3.42 \pm      0.15 &                    ... \\ 
                  3 &       56.11106 &       -43.52301 &          21.300 &    1.911 \pm 0.003 & -1.418 \pm  0.052 &      -2.19^{+0.15}_{-0.15} &                   ... &      -1.85 \pm      0.16 &                    ... \\ 
                  4 &       56.08661 &       -43.54101 &          21.392 &    1.874 \pm 0.003 & -1.518 \pm  0.048 &      -2.52^{+0.15}_{-0.17} &                   ... &      -1.94 \pm      0.20 &                    ... \\ 
                  5 &       56.08898 &       -43.50581 &          21.522 &    1.843 \pm 0.003 & -1.222 \pm  0.047 &      -1.63^{+0.16}_{-0.15} &                   ... &      -1.92 \pm      0.21 &                    ... \\ 
                  6 &       56.08422 &       -43.53636 &          21.578 &    1.860 \pm 0.004 & -1.453 \pm  0.051 &      -2.33^{+0.15}_{-0.16} &                   ... &                      ... &    -1.00 \pm      0.10 \\ 
                  7 &       56.06098 &       -43.52634 &          21.629 &    1.760 \pm 0.003 & -1.384 \pm  0.038 &      -2.19^{+0.12}_{-0.12} &                   ... &      -2.54 \pm      0.21 &                    ... \\ 
                  8 &       56.11060 &       -43.54553 &          21.642 &    1.766 \pm 0.003 & -1.378 \pm  0.035 &      -2.17^{+0.11}_{-0.12} &                   ... &      -1.76 \pm      0.22 &                    ... \\ 
                  9 &       56.07579 &       -43.51999 &          21.772 &    1.640 \pm 0.003 & -1.475 \pm  0.043 &      -2.67^{+0.18}_{-0.21} &                   ... &      -2.63 \pm      0.22 &                    ... \\ 
                 10 &       56.12467 &       -43.54148 &          21.781 &    1.702 \pm 0.004 & -1.389 \pm  0.037 &      -2.25^{+0.12}_{-0.13} &                   ... &                      ... &                    ... \\ 
11\tablenotemark{a} &       56.09614 &       -43.52345 &          21.841 &    1.658 \pm 0.004 & -1.664 \pm  0.050 &      -3.65^{+0.32}_{-0.24} &                <-3.15 &      -2.82 \pm      0.28 &    -2.14 \pm      0.17 \\ 
12\tablenotemark{b} &       56.08598 &       -43.55225 &          21.964 &    1.622 \pm 0.004 & -1.573 \pm  0.045 &      -3.31^{+0.32}_{-0.36} &                   ... &                      ... &                    ... \\ 
                 13 &       56.07677 &       -43.55096 &          22.006 &    1.631 \pm 0.003 & -1.322 \pm  0.053 &      -2.08^{+0.18}_{-0.19} &                   ... &                      ... &                    ... \\ 
                 14 &       56.07042 &       -43.54281 &          22.010 &    1.586 \pm 0.004 & -1.461 \pm  0.045 &      -2.69^{+0.20}_{-0.24} &                   ... &                      ... &                    ... \\ 
                 15 &       56.06578 &       -43.54295 &          22.045 &    1.638 \pm 0.003 & -1.403 \pm  0.061 &      -2.39^{+0.23}_{-0.25} &                   ... &                      ... &                    ... \\ 
                 16 &       56.08129 &       -43.55030 &          22.062 &    1.631 \pm 0.004 & -1.415 \pm  0.058 &      -2.44^{+0.22}_{-0.25} &                   ... &                      ... &                    ... \\ 
                 17 &       56.08408 &       -43.50997 &          22.107 &    1.597 \pm 0.004 & -1.462 \pm  0.047 &      -2.69^{+0.21}_{-0.25} &                   ... &                      ... &                    ... \\ 
                 18 &       56.11464 &       -43.54801 &          22.206 &    1.537 \pm 0.004 & -1.321 \pm  0.035 &      -2.18^{+0.13}_{-0.13} &                   ... &      -2.12 \pm      0.22 &                    ... \\ 
                 19 &       56.07861 &       -43.54173 &          22.241 &    1.602 \pm 0.004 & -1.480 \pm  0.068 &      -2.83^{+0.33}_{-0.47} &                   ... &                      ... &                    ... \\ 
                 20 &       56.11415 &       -43.54755 &          22.336 &    1.565 \pm 0.005 & -1.441 \pm  0.039 &      -2.63^{+0.17}_{-0.19} &                   ... &                      ... &                    ... \\ 
21\tablenotemark{a} &       56.07637 &       -43.53898 &          22.510 &    1.457 \pm 0.005 & -1.597 \pm  0.059 &      -3.66^{+0.34}_{-0.24} &                <-3.05 &                      ... &                    ... \\ 
22\tablenotemark{b} &       56.10017 &       -43.53943 &          22.666 &    1.477 \pm 0.005 & -1.482 \pm  0.045 &      -3.18^{+0.35}_{-0.43} &                   ... &                      ... &                    ... \\ 
                 23 &       56.09489 &       -43.54457 &          22.673 &    1.553 \pm 0.005 & -1.367 \pm  0.061 &      -2.35^{+0.23}_{-0.26} &                   ... &                      ... &                    ... \\ 
                 24 &       56.08373 &       -43.54938 &          22.764 &    1.520 \pm 0.004 & -1.385 \pm  0.080 &      -2.54^{+0.33}_{-0.47} &                   ... &                      ... &                    ... \\ 
                 25 &       56.07893 &       -43.52582 &          22.821 &    1.504 \pm 0.004 & -1.407 \pm  0.069 &      -2.66^{+0.32}_{-0.45} &                   ... &                      ... &                    ... \\ 
                 26 &       56.09459 &       -43.50749 &          22.833 &    1.482 \pm 0.006 & -1.141 \pm  0.065 &      -1.59^{+0.26}_{-0.23} &                   ... &                      ... &                    ... \\ 
                 27 &       56.09084 &       -43.50566 &          22.841 &    1.483 \pm 0.005 & -1.232 \pm  0.076 &      -1.93^{+0.27}_{-0.28} &                   ... &                      ... &                    ... \\ 
                 28 &       56.11173 &       -43.52713 &          22.852 &    1.413 \pm 0.006 & -1.301 \pm  0.090 &      -2.40^{+0.40}_{-0.54} &                   ... &                      ... &                    ... \\ 
                 29 &       56.10807 &       -43.53627 &          22.878 &    1.497 \pm 0.005 & -1.349 \pm  0.074 &      -2.39^{+0.29}_{-0.38} &                   ... &                      ... &                    ... \\ 
30\tablenotemark{b} &       56.09227 &       -43.52335 &          22.923 &    1.480 \pm 0.008 & -1.431 \pm  0.071 &      -2.88^{+0.39}_{-0.56} &                   ... &                      ... &    -1.77 \pm      0.16 \\ 
31\tablenotemark{b} &       56.12278 &       -43.53618 &          22.933 &    1.436 \pm 0.006 & -1.417 \pm  0.055 &      -2.91^{+0.35}_{-0.48} &                   ... &                      ... &                    ... \\ 
                 32 &       56.08301 &       -43.52886 &          22.954 &    1.421 \pm 0.005 & -1.532 \pm  0.095 &      -3.41^{+0.50}_{-0.41} &                <-2.60 &                      ... &    -2.08 \pm      0.21 \\ 
                 33 &       56.09625 &       -43.55913 &          23.075 &    1.452 \pm 0.006 & -1.186 \pm  0.059 &      -1.80^{+0.21}_{-0.20} &                   ... &                      ... &                    ... \\ 
34\tablenotemark{b} &       56.10423 &       -43.52568 &          23.093 &    1.467 \pm 0.007 & -1.416 \pm  0.081 &      -2.87^{+0.43}_{-0.59} &                   ... &                      ... &    -1.88 \pm      0.25 \\ 
                 35 &       56.06941 &       -43.52878 &          23.129 &    1.452 \pm 0.007 & -1.319 \pm  0.070 &      -2.33^{+0.27}_{-0.35} &                   ... &                      ... &                    ... \\ 
                 36 &       56.08063 &       -43.51725 &          23.140 &    1.405 \pm 0.007 & -1.326 \pm  0.071 &      -2.51^{+0.33}_{-0.45} &                   ... &                      ... &                    ... \\ 
37\tablenotemark{b} &       56.06264 &       -43.54250 &          23.152 &    1.432 \pm 0.008 & -1.420 \pm  0.087 &      -2.99^{+0.51}_{-0.62} &                   ... &                      ... &                    ... \\ 
                 38 &       56.07344 &       -43.53673 &          23.161 &    1.409 \pm 0.007 & -1.258 \pm  0.090 &      -2.18^{+0.36}_{-0.46} &                   ... &                      ... &                    ... \\ 
39\tablenotemark{b} &       56.07087 &       -43.53055 &          23.169 &    1.435 \pm 0.007 & -1.369 \pm  0.077 &      -2.66^{+0.39}_{-0.55} &                   ... &                      ... &                    ... \\ 
                 40 &       56.07188 &       -43.53692 &          23.175 &    1.423 \pm 0.007 & -1.269 \pm  0.067 &      -2.16^{+0.26}_{-0.30} &                   ... &                      ... &                    ... \\ 
                 41 &       56.05765 &       -43.52671 &          23.177 &    1.413 \pm 0.008 & -1.271 \pm  0.063 &      -2.19^{+0.24}_{-0.29} &                   ... &                      ... &                    ... \\ 
                 42 &       56.07524 &       -43.52644 &          23.178 &    1.408 \pm 0.006 & -1.359 \pm  0.061 &      -2.65^{+0.32}_{-0.46} &                   ... &                      ... &                    ... \\ 
                 43 &       56.11103 &       -43.55082 &          23.191 &    1.414 \pm 0.007 & -1.256 \pm  0.059 &      -2.12^{+0.22}_{-0.25} &                   ... &                      ... &                    ... \\ 
                 44 &       56.09105 &       -43.53941 &          23.197 &    1.417 \pm 0.007 & -1.294 \pm  0.087 &      -2.33^{+0.35}_{-0.51} &                   ... &                      ... &                    ... \\ 
                 45 &       56.10678 &       -43.54770 &          23.209 &    1.432 \pm 0.007 & -1.223 \pm  0.090 &      -1.97^{+0.33}_{-0.40} &                   ... &                      ... &                    ... \\ 
46\tablenotemark{b} &       56.07859 &       -43.51662 &          23.211 &    1.425 \pm 0.006 & -1.463 \pm  0.057 &      -3.28^{+0.43}_{-0.46} &                   ... &                      ... &                    ... \\ 
47\tablenotemark{b} &       56.11409 &       -43.54579 &          23.218 &    1.376 \pm 0.007 & -1.429 \pm  0.057 &      -3.30^{+0.43}_{-0.45} &                   ... &                      ... &                    ... \\ 
48\tablenotemark{b} &       56.08152 &       -43.53250 &          23.285 &    1.436 \pm 0.009 & -1.392 \pm  0.093 &      -2.84^{+0.50}_{-0.67} &                   ... &                      ... &                    ... \\ 
49\tablenotemark{b} &       56.08677 &       -43.52698 &          23.286 &    1.381 \pm 0.008 & -1.442 \pm  0.074 &      -3.29^{+0.49}_{-0.47} &                   ... &                      ... &    -2.15 \pm      0.22 \\ 
50\tablenotemark{b} &       56.06698 &       -43.53438 &          23.366 &    1.317 \pm 0.007 & -1.363 \pm  0.076 &      -3.15^{+0.53}_{-0.53} &                   ... &                      ... &                    ... \\ 
                 51 &       56.09814 &       -43.54478 &          23.395 &    1.399 \pm 0.006 & -1.155 \pm  0.093 &      -1.76^{+0.36}_{-0.35} &                   ... &                      ... &                    ... \\ 
52\tablenotemark{b} &       56.12947 &       -43.53420 &          23.529 &    1.342 \pm 0.008 & -1.328 \pm  0.096 &      -2.86^{+0.54}_{-0.68} &                   ... &                      ... &                    ... \\ 
                 53 &       56.12470 &       -43.53207 &          23.534 &    1.388 \pm 0.008 & -1.271 \pm  0.077 &      -2.30^{+0.33}_{-0.42} &                   ... &                      ... &                    ... \\ 
                 54 &       56.06552 &       -43.53220 &          23.635 &    1.377 \pm 0.007 & -1.200 \pm  0.099 &      -2.03^{+0.39}_{-0.50} &                   ... &                      ... &                    ... \\ 
                 55 &       56.10222 &       -43.51683 &          23.655 &    1.351 \pm 0.010 & -1.433 \pm  0.083 &      -3.32^{+0.52}_{-0.46} &                <-2.49 &                      ... &                    ... \\ 
                 56 &       56.07829 &       -43.51428 &          23.675 &    1.333 \pm 0.007 & -1.433 \pm  0.073 &      -3.41^{+0.50}_{-0.40} &                <-2.64 &                      ... &                    ... \\ 
57\tablenotemark{b} &       56.11031 &       -43.54366 &          23.875 &    1.271 \pm 0.009 & -1.232 \pm  0.094 &      -2.59^{+0.55}_{-0.74} &                   ... &                      ... &                    ... \\ 
                 58 &       56.11346 &       -43.54091 &          23.903 &    1.328 \pm 0.009 & -1.073 \pm  0.094 &      -1.62^{+0.36}_{-0.37} &                   ... &                      ... &                    ... \\ 
                 59 &       56.11659 &       -43.54202 &          23.987 &    1.280 \pm 0.009 & -1.371 \pm  0.088 &      -3.28^{+0.60}_{-0.49} &                <-2.34 &                      ... &                    ... \\ 
\enddata
\tablenotetext{a}{Stars that we identify in Section \ref{sec:noteworthystars} as being spectroscopically accessible for chemical abundance studies and/or to verify their EMP nature. We discuss them in Section \ref{sec:noteworthystars}}.
\tablenotetext{b}{Stars whose posterior distributions display clear peaks, but which are truncated at the metal-poor end, as discussed in Section \ref{sec:resultsindiv}.}
\end{deluxetable*}
\end{longrotatetable}

\end{document}